\documentclass[acmsmall]{acmart}

\AtBeginDocument{%
  }

\setcopyright{acmlicensed}
\copyrightyear{2018}
\acmYear{2018}
\acmDOI{XXXXXXX.XXXXXXX}

\acmJournal{JACM}
\acmVolume{37}
\acmNumber{4}
\acmArticle{111}
\acmMonth{8}

\usepackage{xcolor}
\usepackage{graphicx}  
\usepackage{subfig} 
\usepackage{dsfont}
\usepackage{hyperref}
\hypersetup{
    colorlinks=true,
} 

\usepackage{amsmath}
\usepackage{url}
\usepackage{enumerate}
\usepackage{multirow}
\usepackage{booktabs}  
\usepackage{makecell}
\usepackage{bm}
\usepackage{pifont}
\usepackage{threeparttable} 
\usepackage{ragged2e}   
\usepackage[edges]{forest}
\usepackage{tikz}
\usepackage{verbatim}
\usepackage{amsthm}
\usepackage{multicol} 

\newtheorem{definition}{Definition}

\definecolor{mygreen}{RGB}{166,219,162} % 
\definecolor{output-black}{RGB}{122,122,122} % 

\begin{document}

\title{Towards Next-Generation LLM-based Recommender Systems: A Survey and Beyond}


\author{Qi Wang}
% \authornote{Both authors contributed equally to this researc.}
\authornote{Equal Contribution.}
\authornote{Corresponding Author.}
\email{qiwang@mails.jlu.edu.cn}
\author{Jindong Li}
\authornotemark[1]
\email{jdli21@mails.jlu.edu.cn}
\author{Shiqi Wang}
\email{shiqiw23@mails.jlu.edu.cn}
\affiliation{%
  \institution{School of Artificial Intelligence, Jilin University}
  \city{Changchun}
  \state{Jilin}
  \country{China}
}

\author{Qianli Xing}
\affiliation{%
  \institution{School of Computer Science and Technology, Jilin University}
  \city{Changchun}
  \state{Jilin}
  \country{China}
}
\email{qianlixing@jlu.edu.cn}

\author{Runliang Niu}
\email{niurl19@mails.jlu.edu.cn}
\author{He Kong}
\email{konghe19@mails.jlu.edu.cn}
\affiliation{%
  \institution{School of Artificial Intelligence, Jilin University}
  \city{Changchun}
  \state{Jilin}
  \country{China}
}

\author{Rui Li}
\affiliation{%
  \institution{MRS, Meta AI}
  \city{Menlo Park}
  \state{California}
  \country{USA}
}
\email{ruili@meta.com}

\author{Guodong Long}
\affiliation{%
  \institution{Australian Artificial Intelligence Institute, University of Technology Sydney}
  \city{Sydney}
  \state{New South Wales}
  \country{Australia}
}
\email{guodong.long@uts.edu.au}

\author{Yi Chang}
\authornotemark[2]
\affiliation{%
  \institution{School of Artificial Intelligence, Jilin University}
  \city{Changchun}
  \state{Jilin}
  \country{China}
}
\email{yichang@jlu.edu.cn}

\author{Chengqi Zhang}
\affiliation{%
  \institution{Department of Data Science and Artificial Intelligence, The Hong Kong Polytechnic University}
  \city{Hong Kong}
  \country{China}
}
\email{chengqi.zhang@polyu.edu.hk}

\renewcommand{\shortauthors}{Qi Wang, Jindong Li, et al.}

\begin{abstract}
Large language models (LLMs) have not only revolutionized the field of natural language processing (NLP) but also have the potential to bring a paradigm shift in many other fields due to their remarkable abilities of language understanding, as well as impressive generalization capabilities and reasoning skills. As a result, recent studies have actively attempted to harness the power of LLMs to improve recommender systems, and it is imperative to thoroughly review the recent advances and challenges of LLM-based recommender systems. Unlike existing work, this survey does not merely analyze the classifications of LLM-based recommendation systems according to the technical framework of LLMs. Instead, it investigates how LLMs can better serve recommendation tasks from the perspective of the recommender system community, thus enhancing the integration of large language models into the research of recommender system and its practical application. In addition, the long-standing gap between academic research and industrial applications related to recommender systems has not been well discussed, especially in the era of large language models. In this review, we introduce a novel taxonomy that originates from the intrinsic essence of recommendation, delving into the application of large language model-based recommendation systems and their industrial implementation. Specifically, we propose a three-tier structure that more accurately reflects the developmental progression of recommendation systems from research to practical implementation, including representing and understanding, scheming and utilizing, and industrial deployment. Furthermore, we discuss critical challenges and opportunities in this emerging field. A more up-to-date version of the papers is maintained at: \url{https://github.com/jindongli-Ai/Next-Generation-LLM-based-Recommender-Systems-Survey}.
\end{abstract}




% \received{xx xx xxxx}
% \received[revised]{xx xx xxxx}
% \received[accepted]{xx xx xxxx}

\maketitle

\section{Introduction}
\label{sec_Introduction}
Recommender systems have become integral to our digital lives. By curating personalized content, products, and services \cite{2023_AI-OPEN_Survey_Information-Retrieval-meets-Large-Language-Models:-A-strategic-report-from-Chinese-IR-community}, these systems facilitate user decision-making processes, enhance user experience, and contribute to the economic success of industries. Recently, large language models (LLMs) have become a cornerstone in enhancing recommender systems due to their powerful generalization capabilities and reasoning skills, which has led to a shift in the realm of recommendation paradigm \cite{2024_arXiv_HSTU_Actions-speak-louder}. More importantly, the emergence of large language models has brought new opportunities to bridge the substantial gap between research and practical deployment, enabling the authentic translation of recommendation algorithms into real-world applications \cite{2024_SIGMOD-Companion_COSMO}. Such advances promise to revolutionize conventional recommendation frameworks and set the stage for a new era of recommendation.

% --------------- table: surveys ---------------
\begin{table*}[b!]
    \centering
    \caption{The comparison between this work and existing surveys.}
    \resizebox{0.93\textwidth}{!}{
    \begin{tabular}{c|c|c|c|c|c|c|c}
    % \hline
    \toprule
        \textbf{Paper} 
        & \textbf{\makecell{Non-\\Generative\\RecSys}} 
        & \textbf{\makecell{Generative\\RecSys}} 
        & \textbf{Scenarios} 
        & \textbf{Academic} 
        & \textbf{Industrial} 
        & \textbf{Pipeline} 
        & \textbf{Hilights} \\ 
        \midrule
        
        \citet{2023_arXiv_Survey_A-Survey-on-Large-Language-Models-for-Recommendation} 
        & \checkmark 
        & \checkmark
        & \makecell{common\\(all kinds)}
        & \checkmark
        &
        & \makecell[l]{(1) Discriminative LLM4REC\\(2) Generative LLM4REC\\Modeling Paradigms:\\(i) LLM Embeddings + RS\\(ii) LLM Tokens + RS\\(iii) LLM as RS}
        & \makecell[l]{focuses on expanding the capacity\\ of language models}\\ \hline
            
        \citet{2023_arXiv_Survey_How-Can-Recommender-Systems-Benefit-from-Large-Language-Models:-A-Survey} 
        & \checkmark
        &  
        & \makecell{common\\(all kinds)}
        & 
        & \checkmark
        & \makecell[l]{(1) Where to adapt to LLM\\(2) How to adapt to LLM}
        & \makecell[l]{from the angle of the whole pipeline\\ in industrial recommender systems} \\ \hline
            
        \citet{2023_arXiv_Survey_A-Survey-on-Language-Models-for-Personalized-and-Explainable-Recommendations} 
        & \checkmark
        &  
        &  \makecell{personalized and\\explainable RecSys}
        & \checkmark
        &
        & \makecell[l]{(1) Explanation Generating \\for Recommendation}
        & \makecell[l]{focuses on utilizing LLMs for \\personalized explanation generating \\task} \\ \hline
            
        \citet{2024_TKDE_Survey_Recommender-Systems-in-the-Era-of-Large-Language-Models-(LLMs)} 
        & \checkmark
        &  
        & \makecell{common\\(all kinds)}
        & \checkmark
        & 
        & \makecell[l]{(1) Pre-training\\(2) Fine-tuning\\(3) Prompting}
        & \makecell[l]{comprehensively reviews such \\domain-specific techniques for \\adapting LLMs to recommendations} \\ \hline
            
        \citet{2024_KDD_Survey_A-Review-of-Modern-Recommender-Systems-Using-Generative-Models-(Gen-RecSys)} 
        &
        & \checkmark 
        & \makecell{(1) interaction\\-driven\\(2) text-driven\\(3) multimodal}
        & \checkmark
        &
        & \makecell[l]{(1) Generative Models for \\Interaction-Driven \\Recommendation\\(2) Large Language Models in \\Recommendation\\(3) Generative Multimodal \\Recommendation Systems}
        & \makecell[l]{aims to connect the key advancements \\in RS using Generative Models \\(Gen-RecSys)} \\ \hline
            
        \citet{2024_arXiv_Survey_Multimodal-Pretraining-Adaptation-and-Generation-for-Recommendation:-A-Survey} 
        & \checkmark
        &  
        & \makecell{multimodal \\recommendation}
        & \checkmark
        &
        & \makecell[l]{(1) Multimodal Pretraining for\\ Recommendation \\(2) Multimodal Adaption for\\ Recommendation \\(3) Multimodal Generation for\\ Recommendation}
        & \makecell[l]{seeks to provide a comprehensive \\exploration of the latest advancements\\ and future trajectories in multimodal\\ pretraining, adaptation, and\\ generation techniques, as well as\\ their applications to recommender\\ systems} \\ \hline
            
        \citet{2024_LREC-COLING_Survey_Large-Language-Models-for-Generative-Recommendation:-A-Survey-and-Visionary-Discussions} 
        &
        & \checkmark 
        & \makecell{common\\(all kinds)}
        & \checkmark
        &
        & \makecell[l]{(1) ID Creation Methods \\(2) How to Do Generative \\Recommendation}
        & \makecell[l]{reviewes the recent progress\\ of LLM-based generative\\ recommendation and provides\\ a general formulation for each\\ generative recommendation\\ task according to relevant research} \\ \hline
        
        \textbf{Ours} 
        & \checkmark 
        & \checkmark 
        % & \makecell{Multimodal\\Sequential\\Explanation\\Conversation\\Cross-Domain\\Personalized} 
        & \makecell{common\\(all kinds)}
        & \checkmark 
        & \checkmark 
        & \makecell[l]{(1) Representing and Understanding\\(2) Scheming and Utilizing\\(3) Industrial Deploying} 
        & \makecell[l]{(1) reviews existing works from \\the 
 perspective of recommender \\system community \\(2) clearly discuss the gap from \\academic research to industrial \\application} \\
        \bottomrule
         
    \end{tabular}
    \label{tab:comparison between this work and existing surveys}
    }
\end{table*}

With increasing efforts to explore large language model (LLM) methods for recommender systems \cite{2024_WWW_CLLM4Rec,2024_SIGIR_IDGenRec,2023_arXiv_Chat-REC,Zhang2023RecommendationAI}, several critical issues merit further investigation.
\textbf{Firstly}, current deep learning methods focus on obtaining the embedding features of users and items via user/item ID to perform similarity calculations for recommendation. However, the precise semantic significance of these embeddings often remains obscure. For instance, a particular embedding dimension may denote a specific attribute, yet such interpretations are seldom understood. Large language models possess two key capabilities that recommender systems lack: (1) they are endowed with a wealth of factual and common sense knowledge, which enables them to provide in-depth details that typically lie beyond the corpus of recommender systems; (2) large language models have the capacity for reasoning, which encompasses the association of items, the analysis of user behavior, and preferences. They are particularly adept at handling complex user behaviors in complex scenarios. Therefore, it is worth discussing how large language models can help recommender systems better represent users/items and enhance the understanding of recommendation behaviors. 

\textbf{Secondly}, existing deep learning-based recommender systems normally follow a pipeline as: data collection, feature engineering, feature encoder, ranking function and recommendation. The emergence of large language models has led to a number of ways to adapt large language models to the existing pipeline. However, the use of large language models should not be limited to this, and their powerful understanding and generation ability can lead to new research paradigms, which are rarely explored in existing works. As a result, this necessitates a comprehensive overview of the potential integration of large language models into recommender systems, which is pivotal for systematically steering both research and practical implementations in the realm of LLM-enhanced recommendation strategies.

\textbf{Last but not least}, the significance of recommender systems lies predominantly in their practical implementation within industrial applications, serving the needs of the industry and its users. While the gap between academia and industry has long been a concern in recommender systems, large language models have brought new opportunities for recommendation algorithms to industrial deployment through their sophisticated natural language processing and understanding abilities. However, large language models exhibit certain limitations in recommendation scenarios, such as the difficulty in assigning precise scores to recommended products, which is crucial for applications such as industrial promotion/advertising. Therefore, it is imperative to comprehensively review the recent advances and challenges of LLM-based recommender systems from the perspective of industry. 

% --------------- figure: Pyramid ---------------
\begin{figure*}[t!]
    \centering
    \includegraphics[width=0.8\linewidth]{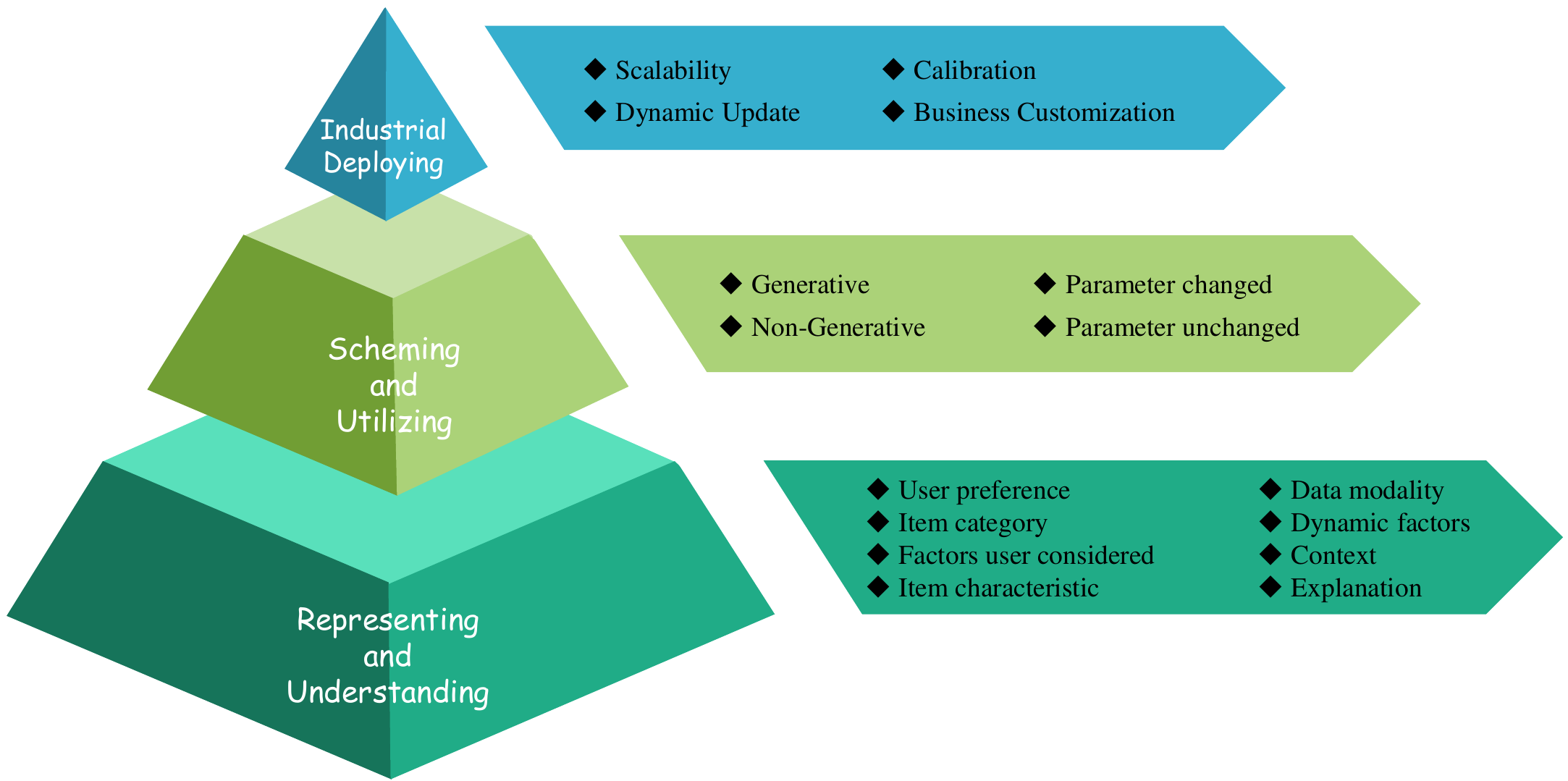}
    \caption{An overview of the proposed three-tier taxonomy for LLM-based recommender systems: (a) representing and understanding, (b) scheming and utilizing, and (c) industrial deploying.}
    \label{fig:fig_1_pyramid}
\end{figure*}

In this survey, we explore the application of Large Language Models (LLMs) in recommender systems, highlighting the limitations of current methodologies and discussing possible challenges and opportunities in the field. It should be noted that our survey is different from some recent surveys on LLM-based recommendation. Specifically, the study explores ways that LLMs can optimize recommendation tasks from the view of the recommender system community, serving the research and implementation of recommender systems in the era of large language models. In contrast, most existing surveys summarize existing work blindly following the taxonomy of LLM techniques from the NLP community \cite{2024_TKDE_Survey_Recommender-Systems-in-the-Era-of-Large-Language-Models-(LLMs), 2023_arXiv_Survey_How-Can-Recommender-Systems-Benefit-from-Large-Language-Models:-A-Survey, 2023_arXiv_Survey_A-Survey-on-Large-Language-Models-for-Recommendation, 2024_arXiv_Survey_Multimodal-Pretraining-Adaptation-and-Generation-for-Recommendation:-A-Survey, 2024_KDD_Survey_A-Review-of-Modern-Recommender-Systems-Using-Generative-Models-(Gen-RecSys), 2024_LREC-COLING_Survey_Large-Language-Models-for-Generative-Recommendation:-A-Survey-and-Visionary-Discussions} or according to recommendation scenarios \cite{2024_LREC-COLING_Survey_Large-Language-Models-for-Generative-Recommendation:-A-Survey-and-Visionary-Discussions, 2023_arXiv_Survey_A-Survey-on-Language-Models-for-Personalized-and-Explainable-Recommendations}. 
As depicted in Fig. \ref{fig:fig_1_pyramid}, \textit{Representing and Understanding} serves the purpose of better \textit{Scheming and Utilizing}, ultimately leading to implementation in the \textit{Industrial Deploying}. More details are shown in Table \ref{tab:comparison between this work and existing surveys}. In summary, this survey makes the following contributions:
\begin{itemize}
    \item We investigate how LLMs can enhance recommendation tasks through the lens of the recommender system community, contrasting with existing surveys that often uncritically adopt the taxonomies of LLM techniques. It would help to offer guidance for next-generation LLM-based recommender systems. 
    \item We propose a three-tier taxonomy to summarize the research on LLM-based recommendation, which improves the representation and understanding of recommender systems, facilitating more effective strategies for their utilization and deployment in industrial settings. To our knowledge, this is the first survey to comprehensively discuss the gap of LLM-based recommender systems from academic research to industrial application, which helps improve the incorporation of large models into the research and practice of the recommender system. 
    \item  We discuss promising challenges and opportunities to explore for the next-generation LLM-based recommender systems, which may help broaden the scope of this under-explored research area.
\end{itemize}

\begin{figure*}[t!]
	\centering
	\resizebox{0.99\textwidth}{!}{
    	\input{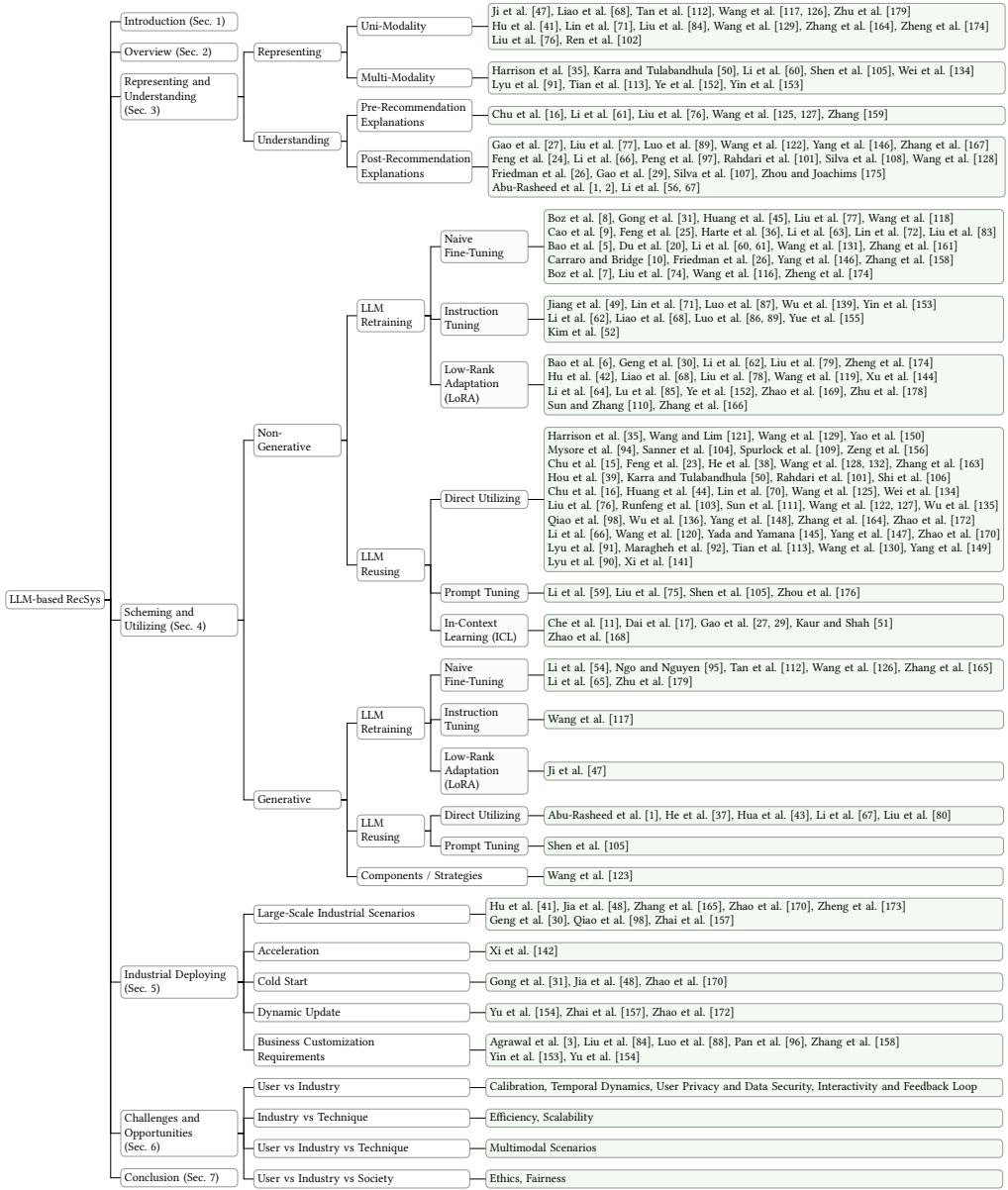}
    	}
	\caption{Structure of this paper with representative works.}
	\label{fig:fig_tree}
\end{figure*}

The remainder of this paper is organized as follows. In Section \ref{sec_Notations and Definitions}, we introduce a brief history of recommender systems, along with the notations and definitions. Section \ref{sec_Representing and Understanding} introduces how LLM can provide a better representation and understanding of recommendation. Section \ref{sec_Scheming and Utilizing} illustrates how to use and scheme LLM in frameworks. Section \ref{sec_Industrial Deploying} discusses how LLM-based recommender systems can bridge the substantial gap between previous research and practical application, facilitating the genuine transition of recommendation algorithms to real-world use. In Section \ref{sec_Challenges and Opportunities}, we highlight the key challenges and opportunities of LLM-based recommender systems. Finally, Section \ref{sec_Conclusion} concludes this survey and draws a hopeful vision for future prospects in research communities of LLM-based recommender systems. The structure of this survey with representative works is illustrated in Fig. \ref{fig:fig_tree}.

% --------------- figure: Structure of Paper ---------------

\section{Overview}
\label{sec_Notations and Definitions}

\subsection{A Brief History of Recommender Systems}

The history of recommender systems is a rich narrative that spans several decades, reflecting the evolution of machine learning techniques. 
The journey began with \textit{simple heuristics} in the early days, where recommendations were often based on the most popular items or those with the highest ratings. These methods were limited in their ability to provide personalized recommendations. 
As data collection became more robust, \textit{collaborative filtering} \cite{2013_IJCAI_TrustMF_Social-Collaborative-Filtering, 2015_AAAI_TrustSVD_Collaborative-Filtering} emerged as a dominant approach. This method exploited the collective preferences of the users to make predictions, either by comparing the similarity between users (user-based collaborative filtering) or between items (item-based collaborative filtering). However, collaborative filtering faced challenges such as the "cold start" problem, where new users or items lacked sufficient data for accurate recommendations. 
During the 2000s, matrix factorization techniques \cite{2009_Computer_MF-for-RecSys, 2007_NeurIPS_PMF, 2010_RecSys_SocialMF} were developed and gained prominence. These methods became an important solution to challenges such as data sparsity. Matrix factorization decomposes the user-item interaction matrix into latent factors, capturing hidden relationships between users and items. This approach provided better scalability and accuracy, particularly in large-scale datasets.
The advent of \textit{neural networks} \cite{2018_IJCAI_DELF_A-Dual-Embedding-based-Deep-Latent-Factor-model} marked a significant shift in the field, allowing more complex and nuanced models. Early neural-based approaches, such as Restricted Boltzmann Machines (RBMs) and Autoencoders, were soon followed by deep learning models \cite{2019_WWW_DANSER, 2020_TKDE_DIffNet++, 2022_SIGIR_DH-HGCN} that could capture non-linear relationships and intricate user behaviors. 
Techniques like Deep Collaborative Filtering \cite{2017_ESWA_Collaborative-Filtering-and-Deep-Learning-based, 2019_RecSys_DSCF_Deep-Social-Collaborative-Filtering} and Neural Collaborative Filtering (NCF) \cite{2019_TOIS_Joint-Neural-Collaborative-Filtering, 2019_SIGIR_NGCF_Neural-Graph-Collaborative-Filtering} extended the capabilities of traditional methods by leveraging the power of multi-layer networks, embedding techniques, and attention mechanisms.

Recently, the integration of \textbf{Large Language Models (LLMs)} has revolutionized recommender systems. LLMs, such as GPT and BERT \cite{2018_arXiv_BERT}, offer an unprecedented ability to understand and generate natural language, which has been used to enhance the performance of recommendation. By processing vast amounts of textual data, LLMs can provide highly personalized, dynamic, and context-sensitive recommendations. They excel at understanding nuanced user preferences, adapting recommendations in real-time, and supporting complex queries that go beyond simple item recommendations. This shift towards LLM-based systems represents a new era in recommendation technology, where the focus is on delivering not just relevant but also contextually and semantically rich experiences to users.

\subsection{Notations and Definitions}
To formally define the process flow of LLM-based recommender systems, in this subsection, we formalize the notations and definitions used for \textit{LLM-based Recommender Systems} as follows.

\subsubsection{Notations} 

%---------------------------- Table: Notations ------------------------------%
\renewcommand{\arraystretch}{0.9}  % to control row spacing
\begin{table}[ht]
\centering
\resizebox{0.7\textwidth}{!}{ % Adjust the width (e.g., 0.8)
    \begin{tabular}{l l | l l}
        \toprule
        Notation & Descriptions & Notation & Descriptions \\
        \midrule
        \( \mathcal{R} \) & user-item interaction data & \( \mathcal{T}_{\text{txt}} \) & text data (e.g., reviews) \\ 
        \( \mathcal{I} \) & image data & \( \mathcal{A} \) & audio data \\
        \( \mathcal{V}_{\text{vid}} \) & video data & \( \mathcal{S} \) & sequence data (e.g., interaction order) \\
        \( \mathcal{T} \) & timestamp data & \( \mathcal{U} \) & the set of users \\
        \( \mathcal{I} \) & the set of items & \( r_{ui} \) & relationship between user \( u \) and item \( i \) \\
        \(\mathbf{x}_u \in \mathbb{R}^d\) & user representation & \(\mathbf{y}_i \in \mathbb{R}^d\) & item representation \\
        \bottomrule
    \end{tabular}
}
\caption{Notations and descriptions.}
\label{tab:notations}
\end{table}

We define the notation used in this paper, which is summarized in Table \ref{tab:notations}. As for $r_{ui}$ in Table \ref{tab:notations}, it can usually be divided into two situations: (1) \textit{Explicit Feedback:} When \(r_{ui}\) represents a rating, it typically takes values from a finite set, such as \(r_{ui}\) being one of \{1, 2, 3, 4, 5\}, where higher values indicate stronger preferences. Occasionally, \(r_{ui}\) might also include a \(0\) to indicate that no rating was given. (2) \textit{Implicit Feedback:} In scenarios involving binary interactions, such as clicks or purchases, \(r_{ui}\) is binary and takes values from \{0, 1\}. Here, \(0\) indicates no interaction, while \(1\) indicates an observed interaction.

Users and items in a recommender system can be represented by feature vectors \(\mathbf{x}_u \in \mathbb{R}^d\) and \(\mathbf{y}_i \in \mathbb{R}^d\), respectively, where \(d\) denotes the feature dimension. These vectors are designed to encapsulate essential characteristics of both users and items. For users, the feature vector \(\mathbf{x}_u\) might include demographic information, behavioral data (e.g., past interactions with items), or inferred preferences. Similarly, the item feature vector \(\mathbf{y}_i\) could encompass attributes such as genre, price, brand, or content descriptors. The dimensionality \(d\) represents the number of features considered, which can be determined based on the specific application or data availability. These vectors serve as the basis for calculating similarity measures or making predictions in various recommendation algorithms.

\subsubsection{Definitions}

\begin{definition}
Recommender Systems.
Recommender systems aim to predict the utility of items for users by leveraging historical interactions. Let \(\mathcal{U} = \{u_1, u_2, \dots, u_m\}\) denote the set of users and \(\mathcal{I} = \{i_1, i_2, \dots, i_n\}\) the set of items. The interaction between user \(u\) and item \(i\) can be represented by a matrix \(\mathbf{R} \in \mathbb{R}^{m \times n}\), where each entry \(r_{ui}\) indicates the interaction level, such as a rating or binary indicator.
\end{definition}

\begin{itemize}
    \item Prediction Function: The recommender system learns a function \(f(u, i | \theta)\) that predicts the interaction \(r_{ui}\) based on parameters \(\theta\). The objective is to minimize the prediction error across all observed interactions:
   \[
   \min_{\theta} \sum_{(u, i) \in \mathcal{D}} \left( r_{ui} - f(u, i | \theta) \right)^2
   \]
   where \(\mathcal{D}\) is the set of observed interactions.

    \item Recommendation: Given the learned function \(f(u, i | \theta)\), the system generates a ranked list of items \(i \in \mathcal{I}\) for each user \(u \in \mathcal{U}\), typically by sorting items based on the predicted scores \(f(u, i | \theta)\).
\end{itemize}

\begin{definition}
LLM-based Recommender Systems.
LLM-based Recommender Systems leverage Large Language Models (LLMs) to enhance or refine recommendation processes. In summary, \textbf{LLM-based Recommender Systems} are defined by their capability to utilize the deep contextual and semantic understanding of LLMs to enhance recommendation tasks. They transform input feature vectors $x_i$ into predictive outputs $\hat{y}_i$ by leveraging the extensive knowledge encoded in the LLM, thereby improving the overall recommendation quality and user experience. The integration of LLMs into recommender systems involves the following components:
\end{definition}

\begin{itemize}
    \item \textbf{Large Language Model (LLM)}: Denoted as $\text{LLM}(\cdot)$, this model is trained on extensive text data and possesses advanced contextual understanding. The LLM is used to process input features and predict recommendations based on deep learning and contextual analysis.
    \item \textbf{Refined Predictive Function} $\hat{f}(\cdot)$: The LLM generates or refines the predictive function. Given the domain $\mathcal{D} = \{\mathcal{X}, P(\mathcal{X})\}$ and task $\mathcal{T} = \{\mathcal{Y}, f(\cdot)\}$, the LLM produces a refined predictive function $\hat{f}(\cdot)$ that maps feature vectors $x_i$ to recommended labels $\hat{y}_i$.
\end{itemize}

\begin{figure*}[t!]
    \centering
    \includegraphics[width=0.99\linewidth]{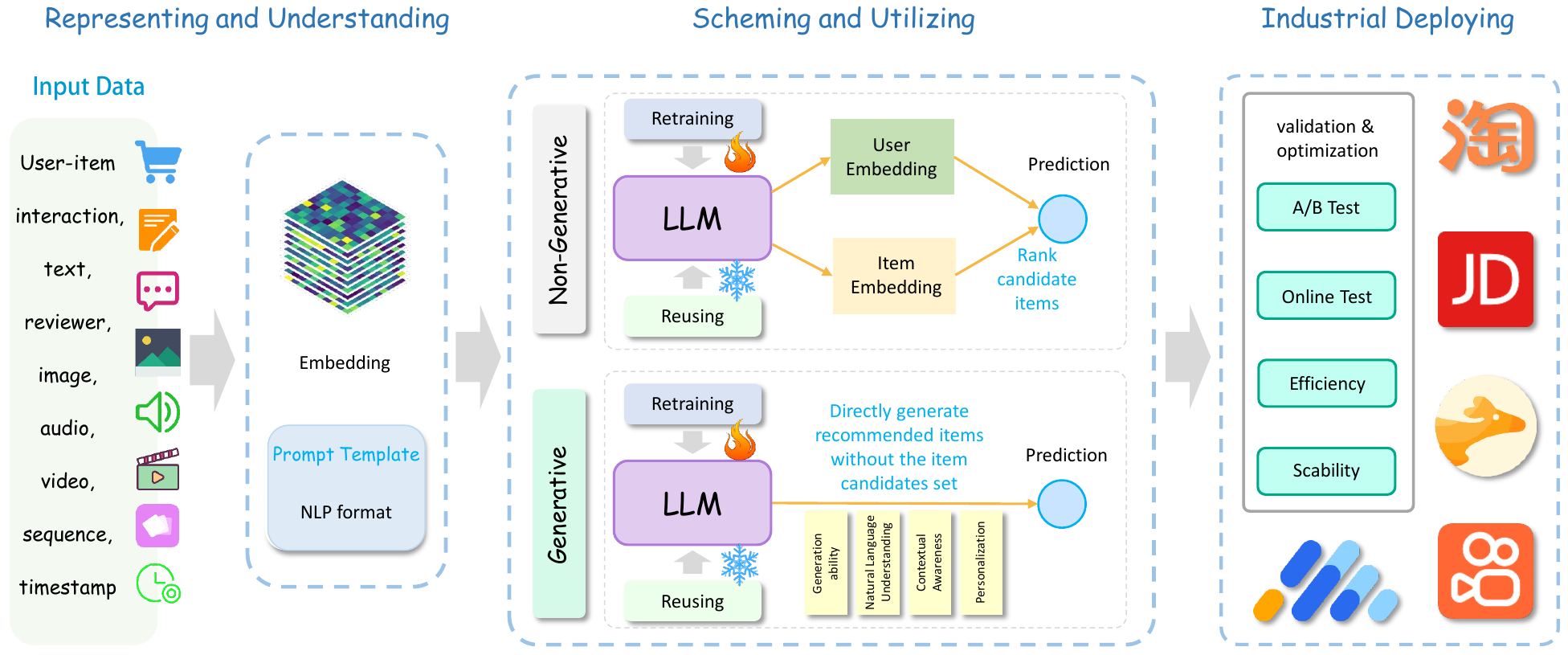}
    \caption{The general pipeline of LLM-based recommender systems.}
    \label{fig:pipeline}
\end{figure*}

The formal representation of the LLM-based recommendation process is:
\[
\hat{f}(x) = \text{LLM}(\mathcal{X}, \mathcal{Y}, \theta)
\]
where $\hat{f}(x)$ is the refined predictive function output by the LLM, $\mathcal{X}$ is the feature space containing user-item interaction vectors, $\mathcal{Y}$ is the label space representing possible recommendations, and $\theta$ represents the model parameters of the LLM. It includes weights and biases learned during training. The general pipeline of LLM-based recommender systems could be illustrated in Fig. \ref{fig:pipeline}.

%------------------ Figure: Pipeline ----------------%

\section{Representing and Understanding}
\label{sec_Representing and Understanding}

With the advent of large language models (LLMs), recommender systems are experiencing a paradigm shift from traditional, closed methodologies to more open and dynamic frameworks. These models bring expansive world knowledge and advanced reasoning capabilities, enhancing the ability of recommender systems to not only generate accurate suggestions but also to improve intermediate processes such as user and item representation. In this evolving landscape, representing and understanding have become pivotal. Representing involves creating nuanced, semantic representations of users and items, categorized into uni-modality and multi-modality approaches. Understanding focuses on elucidating the rationale behind recommendations, with methods that explain either before or after the recommendation is made. Together, these aspects are crucial for developing more effective and transparent recommender systems.

%=================================== Representing START ================================%
\subsection{Representing}
In modern recommender systems, the representation of user-item interactions forms the foundation for generating accurate and personalized recommendations. This representation involves the fusion and processing of multimodal data, including text, image, audio, video, metadata, etc., to capture the full spectrum of user preferences and item characteristics. Let $\mathbf{X}_r$, $\mathbf{X}_{txt}$,$\mathbf{X}_{img}$,$\mathbf{X}_{aud}$, $\mathbf{X}_{vid}$, $\mathbf{X}_{seq}$, $\mathbf{X}_{time}$ denote the extracted features of different modalities including user-item interaction, text data, image data, audio data, video data, sequence data, and timestamp data.

These features are then fused into a unified representation as:
\begin{equation}
\begin{split}
\mathbf{X}_{\text{multi}} = \text{Fusion}(&\mathbf{X}_r, \mathbf{X}_{\text{txt}}, \mathbf{X}_{\text{img}}, \mathbf{X}_{\text{aud}}, \mathbf{X}_{\text{vid}}, \mathbf{X}_{\text{seq}}, \mathbf{X}_{\text{time}})
\end{split}
\end{equation}

While these examples highlight some common types of multimodal data, the spectrum of multimodal inputs is broad and continually evolving, encompassing various other forms of data that may also play crucial roles in enhancing recommender systems.

As recommender systems evolve to meet the diverse needs of users, it becomes crucial to consider both the variety of the data and the complexity of the relationships it captures. This leads us to explore two distinct yet complementary approaches: \textit{Uni-Modality} and \textit{Multi-Modality} recommendations. \textit{Uni-Modality} recommendations focus on leveraging user-item interactions, utilizing graph data to model these interactions and occasionally supplementing it with textual information to refine the understanding of user preferences and item attributes. In contrast, \textit{Multi-Modality} approaches extend beyond the boundaries of uni-modality by integrating data from multiple sources or modalities, such as text, image, and video. This expansion leverages the availability of diverse data sources to provide a richer understanding of user preferences and item attributes, offering a more comprehensive and flexible recommendation strategy. By distinguishing between these approaches, we can better understand how each contributes to the overall goal of providing accurate and personalized recommendations.

\subsubsection{Uni-Modality}

In uni-modality recommendations, the primary approach involves leveraging graph-based methods to model user-item interactions. These methods emphasize the structural relationships between entities, capturing patterns within the data to predict user preferences. To enhance the effectiveness of the recommender system, textual information is occasionally integrated into the graph structure, adding semantic context that enriches the understanding of user preferences and item attributes. This combination of structural and semantic data allows for more accurate and personalized recommendations, particularly in scenarios like sequential and session-based recommendations, where the dynamic and evolving nature of user interactions requires real-time adaptability and responsiveness.

\begin{definition}
Uni-Modality Recommendation.  
Uni-Modality refers to recommender systems that leverage data from a single modality. In this approach, user-item interactions \( \mathcal{R} \) are represented primarily through one type of data, such as user ratings or interaction histories, without incorporating other data sources like image, audio, or video. However, in some cases, textual information \( \mathcal{T}_{\text{txt}} \) is integrated to enhance the representation within this single modality.
\end{definition}

\cite{2023_arXiv_LLaRA_Large-Language-Recommendation-Assistant} contends that using only ID-based or text-based representations of item sequences to prompt large language models does not fully leverage their potential for sequential recommendation. It proposes that LLMs need to develop a deeper understanding of the behavioral patterns embedded in the sequential interactions. This work investigates the alignment between LLMs and sequential recommender systems, moving beyond the use of simple ID-based or text-based prompting. It proposes treating the "sequential behaviors of users" as a new modality for LLMs in recommender systems and aligning it with the language space.

\cite{2023_arXiv_DRDT_DRDT:Dynamic-Reflection-with-Divergent} focuses on sequential recommendation and proposes a novel reasoning principle: Dynamic Reflection with Divergent Thinking within a retriever-reranker framework. 
\cite{2024_WWW_SINGLE_Modeling-User-Viewing-Flow} proposes a user viewing flow modeling (SINGLE) method to recommend articles for users. It models the constant and instant viewing flows to better represent user interests, enhancing the extraction of both general and real-time preferences for recommendations.
\cite{2024_WWW_LLM-TRSR_Harnessing} introduces an innovative framework for utilizing Large Language Models in text-rich sequential recommendation (LLM-TRSR). The method involves several key steps: it begins by extracting the user's behavioral history sequence and converting it into an extended text format. This text is then segmented into multiple blocks, with each block being designed to fit within the processing limits of large language models. Finally, an LLM-based summarizer is employed, which synthesizes these blocks to produce a comprehensive summary of user preferences.
\cite{2024_SIGIR_FineRec_FineRec:Exploring-Fine-grained-Sequential-Recommendation} introduces a new perspective to handle sequential recommendation, where the attribute-opinions of reviews are explored to finely reveal user preferences and item characteristics. To represent users/items under distinct attributes, this work creates a unique attribute-specific user-opinion-item graph for each attribute.
\cite{2024_WWW_SAID_Enhancing-sequential-recommendation} focuses on sequential recommendation and introduces SAID, a framework designed to leverage LLMs for learning semantically aligned item ID embeddings directly from text. For each item, SAID uses a projector module to convert an item ID into an embedding vector, which is then processed by an LLM to generate the precise descriptive text tokens associated with the item. The resulting item embeddings are optimized to capture the detailed semantic information contained in the textual descriptions.
\cite{2023_arXiv_TransRec_a_multi} introduces a novel multi-facet paradigm, called TransRec, to connect LLMs with recommendation systems. Specifically, TransRec utilizes multi-facet identifiers, integrating ID, title, and attributes to capture both uniqueness and semantic meaning.
\cite{2023_arXiv_SAGCN_Understanding-Before-Recommendation} introduces a chain-based prompting technique designed to uncover semantic aspect-aware interactions, providing more detailed insights into user behavior at a fine-grained semantic level. To effectively leverage the extensive interactions across different aspects, the work proposes the semantic aspect-based graph convolution network (SAGCN), a straightforward yet powerful method. 
\cite{2024_WWW_RLMRec_Representation-Learning-with} introduces a collaborative profile generation paradigm and a reasoning-driven system prompt, emphasizing the integration of reasoning processes in the generated output. RLMRec utilizes contrastive and generative alignment techniques to align collaborative filtering (CF) relational embeddings with large language model semantic representations, effectively mitigating feature noise.

\cite{2024_arXiv_Leveraging-Edge-Information} addresses the challenge of fully underutilized edge information in LLMs, particularly within the key attention mechanism. The proposed method enhances the model's understanding by incorporating the direct relationship between users and items and constructing second-order relationships between items, providing a complex association absent in traditional recommendation data.
\cite{2024_ECIR_GenRec} leverages the detailed information inherently present in item names, which often include features amenable to semantic analysis, thereby facilitating a better understanding of an item’s potential relevance to the user. The proposed GenRec model enhances the generative recommendation performance by incorporating textual information.
\cite{2024_SIGIR_IDGenRec} addresses the item encoding problem in generative recommendation systems by introducing IDGenRec which incorporates textual ID learning. It proposes an ID generator to produce unique, concise, and semantically rich textual IDs that are platform-agnostic and based on human vocabulary. 
\cite{2024_WWW_CLLM4Rec} extends the vocabulary of pre-trained LLMs user/item ID tokens to faithfully model user/item collaborative and content semantics. Additionally, the work proposes a novel "soft+hard prompting" strategy to effectively get user/item collaborative/content token embeddings via language modeling on recommendation system-specific corpora. It splits each document into a prompt which consists of heterogeneous soft (user/item) tokens and hard (vocab) tokens and a main text which consists of homogeneous item tokens or vocab tokens.
\cite{2023_arXiv_Multiple-KV} considers multiple key-value data, because this common scenario is prevalent in real-world applications, where user information (e.g., age, occupation) and item details (e.g., title, category) contain multiple keys.

\subsubsection{Multi-Modality}

In multi-modality recommendations, leveraging various types of data—such as text, images, audio, video, and metadata—enhances the recommender system's capability to provide more relevant suggestions. Techniques like graph augmentation and textual information integration are crucial in this context. By enriching user-item interactions and incorporating diverse attributes, these methods tackle issues like sparse feedback and low-quality side information. The use of multimodal large language models (MLLMs) further supports the integration of complex, real-time data sources, such as screenshots of user activities. This approach emphasizes interpretability, robustness, and adaptability across different domains and modalities, ultimately improving the effectiveness of recommender systems.

\begin{definition}
Multi-Modality Recommendation.  
Multi-Modality refers to recommender systems that integrate data from multiple modalities, represented as \( \mathcal{D} = \{(\mathcal{R}, \mathcal{T}_{\text{txt}}, \mathcal{I}, \mathcal{A}, \mathcal{V}_{\text{vid}}, \mathcal{S}, \mathcal{T})\} \), to provide a richer and more comprehensive representation of both users and items. By combining information from text, image, audio, video, sequence, and timestamp, the system is able to capture diverse aspects of user preferences and item characteristics, leading to more robust and nuanced recommendations.
\end{definition}

\cite{2024_WSDM_LLMRec} leverages LLMs to augment graphs in recommender systems by enhancing user-item interaction edges, item node attributes, and user node profiles. It tackles the scarcity of implicit feedback signals by empowering LLMs to explicitly reason about user-item interaction patterns. Furthermore, it overcomes the issue of low-quality side information by generating user and item attributes and implementing a denoised augmentation mechanism. 
\cite{2024_PAKDD_InteraRec} proposes an innovative approach that utilizes information from screenshots of users’ internet browsing activities. This method leverages MLLMs, which are skilled at processing and generating content across various modalities. By using screenshots instead of weblogs, the system gains improved interpretability. The visual nature of screenshots provides a clear and transparent depiction of user actions, significantly enhancing the understanding of the LLM's inferences. 
\cite{2023_ICDM-Workshop_Zero-shot-Recommendation} presents a method for zero-shot recommendation of multimodal non-stationary content, leveraging recent advancements in generative AI. The proposed approach involves rendering inputs of different modalities as textual descriptions and utilizing pre-trained LLMs to obtain their numerical representations by computing semantic embeddings.
\cite{2024_arXiv_LLM4POI_large} focuses on enhancing recommendations in location-based social networks (LBSNs) and introduces an innovative LLM4POI framework for the next point-of-interest (POI) recommendation. This framework differs from conventional numerical approaches, which typically require data transformation and the use of various embedding layers. Instead, LLM4POI retains the original format of heterogeneous location-based social network data, thereby preserving the contextual information without any loss.
\cite{2023_RecSys_HKFR_hetero} proposes a new paradigm for user modeling that extracts and integrates diverse knowledge from heterogeneous user behavior data. By transforming structured user behavior into unstructured heterogeneous knowledge, it effectively captures user interests. In the context of Meituan Waimai, user heterogeneous behavior includes: multiple behavior subjects, such as merchants and products; multiple behavior contents, such as exposure, clicks, and orders; multiple behavior scenarios, such as APP homepage and mini-programs.
\cite{2024_arXiv_MMREC-LLM-Based-Multi-Modal-Recommender} leverages the reasoning and summarization capabilities of LLMs to effectively process multi-modal information. This study utilizes LLMs to summarize user review texts, capturing subtle user behaviors and preferences, and to generate descriptive text for images, extracting implicit insights about businesses and products.
\cite{2024_arXiv_MLLM-MSR_Harnessing-Multimodal-Large} introduces a novel image summarization method based on Multimodal Large Language Models (MLLMs) to recurrently summarize user preferences across multiple modalities. This approach enables a deeper understanding of user interactions and interests over time, enhancing the ability to capture and predict user behavior. Specifically, this work first utilizes an MLLM-based item summarizer to extract image features from a given item and convert the image into text, enabling more effective integration of visual information into the recommendation process.
\cite{2024_arXiv_X-Reflect_Cross-Reflection-Prompting-for-Multimodal-Recommendation} proposes a novel approach called cross-reflection prompting, referred to as X-REFLECT. The core idea behind cross-reflection prompting is to supply large multimodal models (LMMs) with both textual and visual context while explicitly guiding the models to determine whether these pieces of information are supportive or contradictory. The outputs from the LMMs are then transformed into embeddings, which are used as input item embeddings for the subsequent recommendation module.

\cite{2024_WWW_PMG_Personalized-Multimodal-Generation} takes into account two categories of user behaviors: historical clicks and conversations. The input features are multimodal, encompassing text, images, audio, and more. It utilizes an LLM to preprocess and summarize the text features of each item and conversation into concise sentences. For other modalities, it converts them into text using a captioning model (e.g., BLIP-2 \cite{2023_arXiv_BLIP-2}, CLAP \cite{2023_ICASSP_CLAP}) or a multimodal LLM (e.g., MiniGPT-4 \cite{2023_arXiv_MiniGPT-4}, mPLUG-Owl \cite{2023_arXiv_mPLUG-Owl}) that can handle various types of input. The goal of this preprocessing step is to condense the features, eliminate redundancy, and maintain long-term context.
%=================================== Representing END ================================%

%=================================== Understanding-Explanation START ================================%
\subsection{Understanding}

Recommender systems based on large language models leverage external world knowledge and advanced reasoning capabilities. These models can incorporate external knowledge, particularly information specific to users and items like user preferences, item attributes, and behavioral patterns. Furthermore, due to their strong reasoning abilities, LLMs can offer deep insights into user motivations, as well as relationships between users, items, and their broader social context. Consequently, they enable a more profound understanding of the underlying rationale behind recommendations. Current methods for explanation in recommender systems can be broadly categorized based on the stage of the recommendation process: pre-recommendation explanations and post-recommendation explanations. The specific definitions and relevant works are introduced below.

\subsubsection{Pre-Recommendation Explanations}

\begin{definition}
Pre-Recommendation Explanations. 
This paradigm emphasizes generating explanations for items prior to the recommendation process, offering a clear rationale for why each item is being considered before the final decision is made. It includes (i) utilizing reasoning graphs, (ii) leveraging known relationships between nodes (such as user-item interactions, item similarities, and social connections), (iii) utilizing LLM generates transparent and interpretable intermediate insights through multi-source information extraction and reasoning integration to justify why certain items are considered for recommendation. Let \( \mathbf{h}_u \) denote the latent representation of user \( u \), and \( \mathbf{h}_i \) the latent representation of item \( i \). Based on this, the system computes a score \( s(u, i) = f(\mathbf{h}_u, \mathbf{h}_i, \mathcal{E}_i) \) that incorporates the explanation into the recommendation process:
\[
\mathcal{E}_i = \text{Explain}(\mathbf{h}_u, \mathbf{h}_i)
\]
The function \( \text{Explain}(\cdot) \) generates the explanation before ranking the items, providing insights into why an item is likely a good match.

In this context, \( f(\mathbf{h}_u, \mathbf{h}_i, \mathcal{E}_i) \) is a function that computes a relevance score by combining the user's latent representation, the item's latent representation, and the explanation \( \mathcal{E}_i \), which justifies why a particular item is being considered for recommendation.
\end{definition}

\cite{2023_arXiv_SAGCN_Understanding-Before-Recommendation} leverages the large model's ability to extract semantics from raw comments and proposes using a chain-based prompting approach to obtain semantic aspect-aware reviews from user-item reviews, which are used to discern aspect-aware interactions. The embeddings are enriched by learning from different aspects, and interpretability is improved by integrating semantic aspects.
\cite{2024_arXiv_SKarREC_Learning-Structure-and-Knowledge} focuses on concept recommendation. It utilizes the structural relationships between concepts to help LLMs generate their explanations, which resolves the ambiguity problem in concept explanations. Subsequently, they use the names, explanations, and preceding and succeeding nodes as the textual descriptions for subsequent concept recommendations.
\cite{2024_arXiv_LLMHG_Llm-guided-multiview-hypergrap-learning} proposes a plug-and-play recommendation enhancement framework, LLMHG, which synergizes the reasoning capabilities of LLMs with the structural advantages of hypergraph neural networks. By effectively analyzing and interpreting the subtle nuances of individual user interests, this framework pioneers in enhancing the interpretability of recommender systems.
\cite{2024_arXiv_RDRec_Rationale-Distillation} leverages LLMs through rationale distillation to extract user preferences and item attributes from reviews, obtaining clear textual knowledge. This knowledge is then applied to the task of explanation generation for a user’s interactions.
\cite{2023_RecSys_CPR_User-centric-conversational-recommendation} proposes a graph-based conversational path reasoning (CPR) framework that represents conversations as interactive reasoning over a user-item-attribute graph. This allows capturing user preferences and explaining recommendations based on relationships in the graph. 
\cite{2024_AAAI_LLMRG_Improving-Recommendations-through-LargeLanguageModel} adopts a prompt-based framework that utilizes LLMs to generate new reasoning chains. The prompt takes the next item, existing reasoning chains, and user attributes as input. It outputs a comprehensive set of potential new reasoning chains to explain why a user might choose the next item.

\subsubsection{Post-Recommendation Explanations}

\begin{definition}
Post-Recommendation Explanations.  
In this paradigm, explanations are provided after the recommendation is made, allowing users to understand why certain items were suggested. Given a set of recommended items $ \mathcal{I}_u^* = \{i_1^*, i_2^*, \dots, i_k^*\} $, the system generates a corresponding set of explanations $ \mathcal{E}_u^* = \{\mathcal{E}_{i_1^*}, \mathcal{E}_{i_2^*}, \dots, \mathcal{E}_{i_k^*}\} $. Here, $ \mathcal{E}_{i_j^*} $ represents the explanation for why item $ i_j^* $ was included in the recommendations for user $ u $. The challenge lies in ensuring that these explanations are both informative and personalized, aligning with the user's preferences and the specific context of the recommendation.  
Let \( s(u, i_j^*) \) denote the relevance score between user \( u \) and item \( i_j^* \), which reflects how well the item fits the user's preferences. Mathematically, the problem can be expressed as:
\begin{equation}
\mathcal{E}_{i_j^*} = \text{Explain}(s(u, i_j^*), \mathbf{h}_u, \mathbf{h}_{i_j^*})
\end{equation}
where \( \text{Explain}(\cdot) \) is a function that generates the explanation based on the recommendation score \( s(u, i_j^*) \) and the latent features of the user \( \mathbf{h}_u \) and item \( \mathbf{h}_{i_j^*} \).
\end{definition}

\cite{2023_arXiv_LLMXRec_Unlocking-the-potential} proposes LLMXRec, a two-stage interpretable recommendation framework aimed at further enhancing explanation quality through the use of LLMs. It emphasizes the close collaboration between the recommendation model and the LLM-based explanation generator.
\cite{2024_AAAI_LLM2ER-EQR_Fine-Tuning-Large} designs two explainable quality reward models to fine-tune the backbone model within a reinforcement learning paradigm, enabling it to generate high-quality explanations.
\cite{2023_arXiv_LLM4Vis_Explainable-visualization-recommendation} proposes LLM4Vis, a ChatGPT-based prompting method that performs visualization recommendations and returns human-like explanations. It iteratively improves the generated explanations by considering previous generations and template-based prompts.
\cite{2024_arXiv_ChatGPT-based-CRS_Navigating-User-Experience} develops a ChatGPT-based conversational recommendation systems to study the impact of prompt guidance (PG) and recommendation domain (RD) on overall user experience, finding that PG significantly enhances system explainability, adaptability, perceived ease of use, and transparency.
\cite{2024_arXiv_DRE_Generating-Recommendation-Explanations} introduces data-level recommendation explanation (DRE), a non-intrusive explanation framework for black-box recommendation models, which employs a data-level alignment method to align the explanation module with the recommendation model.
\cite{2024_arXiv_Uncertainty-Aware-Explainable} develops a model using the ID vectors of user and item inputs as prompts for GPT-2, optimized through a joint training mechanism within a multi-task learning framework for both the recommendation and explanation tasks.
\cite{2024_arXiv_Where-to-Move-Next} focuses on the next point-of-interest (POI) recommendation task, by considering important factors to prompt the LLM to recommend the Top-K POIs and provide explanations for the returned suggestions.
\cite{2024_IUI_Leveraging-ChatGPT-for-Automated} uses ChatGPT's conversational abilities to provide human-like recommendation explanations, the user’s perceptions of the explanations generated by ChatGPT were evaluated.
\cite{2024_WSDM_Logic-Scaffolding_Personalized-aspect-instructed} proposes a framework called Logical Scaffolding, which combines aspect-based explanations with the concept of chain-of-thought prompting to generate explanations through intermediate reasoning steps.
\cite{2023_arXiv_Leveraging-Large-Language-Models-in-Conversational-Recommender} employs a joint ranking/explanation module, which uses LLMs to extract user preferences from ongoing conversations. It generates natural language justifications for each item displayed to the user, enhancing the system's transparency.
\cite{2023_arXiv_Chat-rec_Towards-interactive-and-explainable} proposes a new paradigm called Chat-Rec, which transforms user profiles and historical interactions into prompts. It utilizes ChatGPT to learn user preferences, making recommendations more interpretable.
\cite{2023_IntRS_Leveraging-Large-Language-Models} develops a web application that leverages ChatGPT to generate movie recommendations and explanations of said recommendations based on user preferences.
\cite{2023_arXiv_GPT-as-a-Baseline} establishes that modern LLMs are a promising source of post-hoc explanations that could accompany item recommendations with relevant summaries to improve user satisfaction.
\cite{2024_arXiv_Supporting-student-decisions} leverages the potential of knowledge graphs (KGs) for modeling educational content to support LLM-based chatbots in generating more relevant explanations for learning recommendations.
\cite{2024_arXiv_Knowledge-Graphs-as-Context} combines knowledge graphs and GPT models, KGs are utilized as a source of contextual information to support LLMs in generating more relevant explanations for learning recommendations.
\cite{2023_Electronics_Bookgpt_A-general-framework} proposes a book recommendation system framework, BooKGPT. It can even perform personalized interpretable content recommendations based on readers' attributes and identity information.

%----------------- Figure: Differetn kinds of RecSys ---------------------%
\begin{figure*}[t!]
    \centering
    \includegraphics[width=0.95\linewidth]{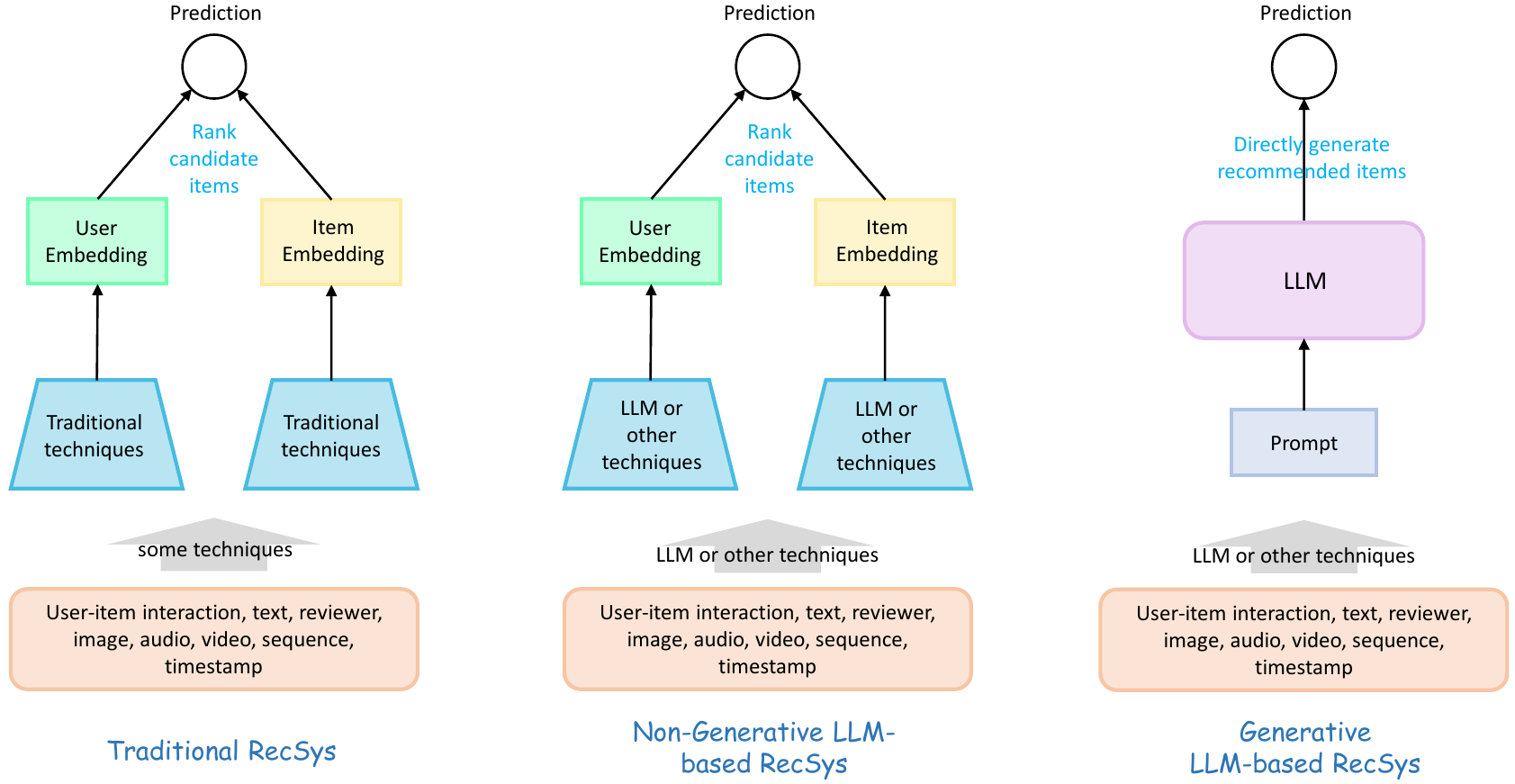}
    \caption{Different paradigms for recommender systems: (a) Traditional Recommender System; (b) Non-Generative LLM-based Recommender System; and (c) Generative LLM-based Recommender System.}
    \label{fig:different_paradigm_of_RecSys}
\end{figure*}
%-------------------------------------------------------------------------%

\cite{2024_AAAI_LLMRec_Benchmarking-Large-Language-Models} designs LLMRec, an LLM-based recommendation system designed for benchmarking LLMs, is developed and found that supervised fine-tuning (SFT) can enhance LLMs' capabilities in the task of explanation generation.
\cite{2024_arXiv_PAP-REC_Personalized-Automatic-Prompt} proposes a framework for generating personalized automatic prompts for recommendation language models, PAP-REC, which replaces manual prompt design and can handle the task of explanation generation.
\cite{2024_NAACL_RecMind_Large-Language-Model-Powered} designs RecMind, an LLM-powered autonomous recommender agent with a self-inspiring algorithm that improves explanation generation by leveraging external knowledge and historical information.
\cite{2023_CIKM_POD_Prompt-distillation-for-efficient} extracts discrete prompts into a set of continuous prompt vectors, which are used as prompts in interpretable tasks.
%=================================== Understanding END ================================%
\section{Scheming and Utilizing}
\label{sec_Scheming and Utilizing}

The emergence of LLMs has introduced a new paradigm in recommender systems, sparking extensive research into how to effectively integrate LLMs into recommendation frameworks. The research in this area can be categorized into \textit{Non-Generative LLM-based} and \textit{Generative LLM-based Approaches}, depending on whether the framework requires calculating rating scores for each candidate to determine recommendations. The distinctions between these approaches and traditional recommender systems are illustrated in Fig. \ref{fig:different_paradigm_of_RecSys}. To provide a clearer understanding of how LLMs are employed within these frameworks, Fig. \ref{fig:LLM-using} presents a further classification of the approaches into two main strategies: \textit{LLM Retraining} and \textit{LLM Reusing}. This classification is based on whether the parameters of the LLMs are changed or unchanged.

To facilitate the discussion in the following subsections, we provide several definitions related to these strategies:

%----------------------------- Figure: LLM-using --------------------------%
\begin{figure*}[t!]
    \centering
    \includegraphics[width=0.95\linewidth]{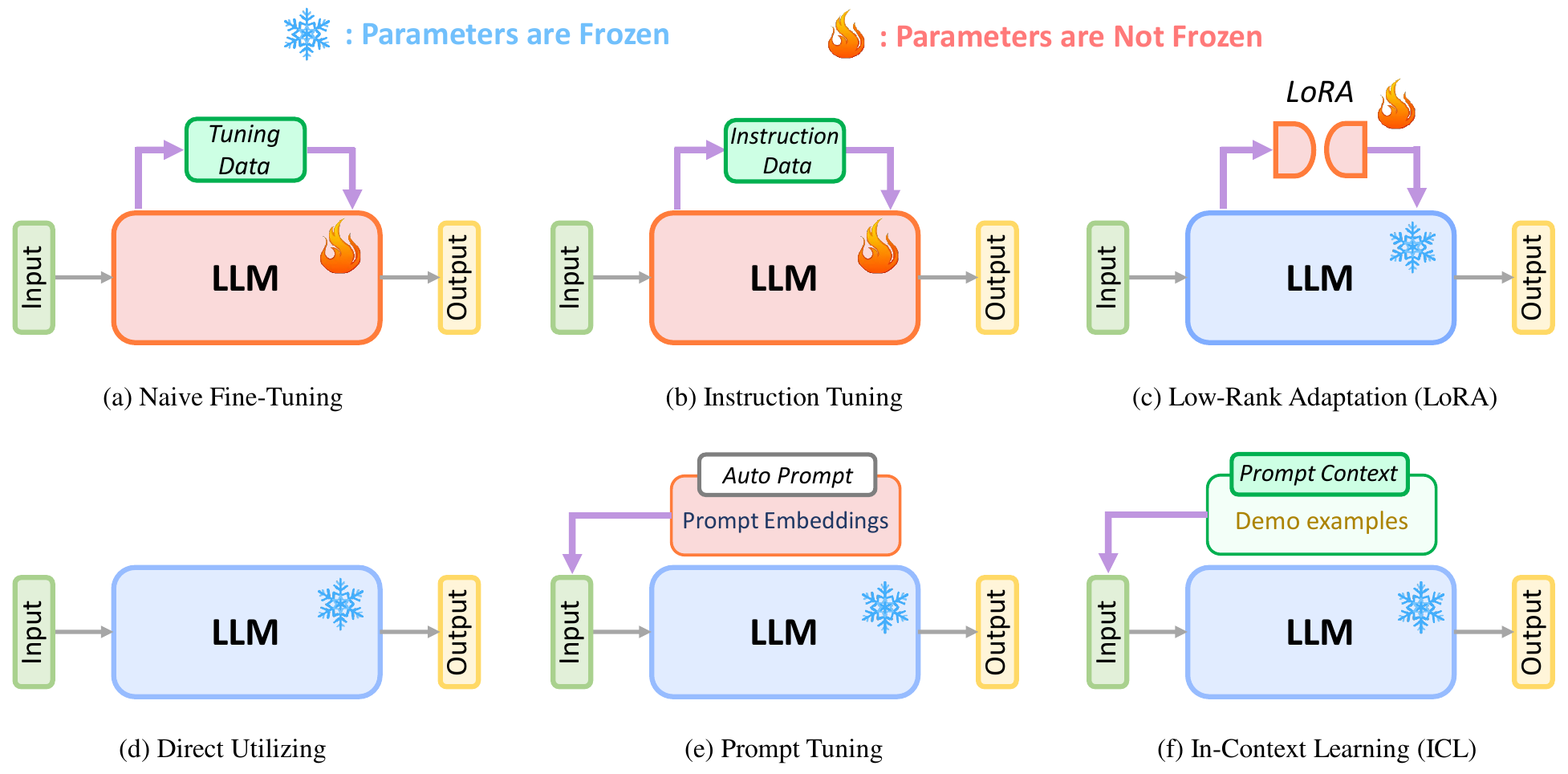}
    \caption{Different paradigms for LLM-based recommender systems: (a) \textit{Naive Fine-Tuning}: tailors the model to excel in specific recommendation tasks using domain/task-specific data; (b) \textit{Instruction Tuning} \cite{2024_arXiv_Survey_instruction_tuning}: optimizes the model's ability to follow diverse recommendation instructions and queries; (c) \textit{Low-Rank Adaptation (LoRA)} \cite{2021_ICML_LoRA_low_rank}: efficiently adjusts recommendation models with minimal changes, keeping the core parameters intact; (d) \textit{Direct Utilizing}: utilizes the model directly in its pre-trained state without additional fine-tuning or training; (e) \textit{Prompt Tuning}: optimizes specific recommendation prompts to tailor the model's responses, while keeping the underlying model parameters unchanged; and (f) \textit{In-Context Learning (ICL)}: uses contextual examples and cues within the input to guide the model's recommendations dynamically, without altering the model's parameters.}
    \label{fig:LLM-using}
\end{figure*}
%-------------------------------------------------------------------------%

\begin{definition}
Non-Generative LLM-based Recommendation.
Non-Generative LLM-based Recommendation is a paradigm that utilizes large language models to enhance traditional recommendation tasks by incorporating the LLMs' understanding of natural language into the recommendation process. Unlike generative approaches, non-generative LLM-based methods do not generate recommendations directly as natural language outputs. Instead, they employ LLMs to improve the accuracy and relevance of the recommendation models, such as by leveraging the semantic understanding embedded in pre-trained language models to enhance ranking, scoring, or feature extraction. These methods typically involve multi-stage processes where the LLMs contribute to specific stages, such as feature enrichment, or ranking.
\end{definition}

\begin{definition}
Generative LLM-based Recommendation.
Generative LLM-based Recommendation is a paradigm that leverages large language models to perform recommendation tasks by transforming them into natural language tasks. This approach allows for generative recommendations, where the system directly produces the items to recommend instead of computing a ranking score for each candidate item as seen in traditional recommendation models. This method often streamlines the recommendation process from multi-stage filtering to a single-stage generation \cite{2023_SIGIR-AP_How-to-Index-Item-IDs, 2024_ECIR_GenRec, 2024_ProLLM4Rec_Prompting-Large-Language-Models, 2023_arXiv_Survey_Large-Language-Models-for-Generative-Recommendation:A-Survey-and-Visionary-Discussions}.
\end{definition}

\begin{definition}
LLM Retraining.  
LLM Retraining refers to the process of modifying the parameters \( \text{LLM}_{\theta} \) of a pre-trained large language model to adapt it to a specific recommendation task. This can involve techniques such as fine-tuning, where the LLM's knowledge is aligned with the particular characteristics and data of the recommender system. The goal of retraining is to enhance the personalization, accuracy, and effectiveness of the recommendations by tailoring the LLM’s capabilities to the domain-specific requirements of the task. This can be mathematically represented as optimizing a loss function \( \mathcal{L}(\text{LLM}_{\theta}; \mathcal{D}_{\text{task}}) \), where \( \mathcal{D}_{\text{task}} \) denotes the dataset specific to the recommendation task.
\end{definition}

\begin{definition}
LLM Reusing.  
LLM Reusing involves utilizing a pre-trained large language model without or with only minimal modifications to its parameters. This strategy capitalizes on the LLM's pre-existing capabilities and knowledge, focusing on optimizing how the model is employed within the recommender system. Methods under this category typically involve adapting how the inputs \( \mathcal{X} \), outputs \( \hat{y} \), or intermediate processing stages are handled, without altering the LLM’s core parameters \( \text{LLM}_{\theta} \). LLM Reusing is particularly advantageous for maintaining computational efficiency and preserving the generalization power of the original model while still improving recommendation performance.
\end{definition}

This classification framework enables a thorough understanding of different LLM-based recommendation strategies. By clearly distinguishing between these approaches, we can better assess and select the most appropriate strategy for specific recommendation scenarios. This comprehensive understanding ultimately facilitates the effective and efficient application of LLMs in recommender systems, ensuring that their capabilities are leveraged in a manner that aligns with both the requirements and constraints of the task.

\subsection{Non-Generative LLM-based Approaches}
Non-generative approaches typically require calculating the ranking score for each candidate individually to determine the recommendation results. We categorize these non-generative methods into LLM retraining and LLM reusing. LLM retraining involves modifying the parameters of the LLM, whereas LLM reusing requires no or only minimal parameter changes.

The Non-Generative approach focuses on utilizing the pre-trained knowledge of an LLM to recommend items or actions without creating new content. Instead of generating novel outputs, it draws directly from the model’s learned representations to make predictions or suggestions based on existing data patterns. Let \( \mathbf{X} = \{x_1, x_2, \dots, x_n\} \) represent the input data, where each \( x_i \) is a feature vector corresponding to a user or context, and \( \mathbf{Z} = \{z_1, z_2, \dots, z_k\} \) represent a set of candidate items or actions that can be recommended.

The Non-Generative approach involves utilizing a pre-trained large language model to rank or select items from \( \mathbf{Z} \), based on the input data \( \mathbf{X} \), without generating new items. This is achieved through a function \( R_\phi(\cdot) \), parameterized by \( \phi \), which calculates relevance scores for each candidate item given the input data. The set of relevance scores \( \mathbf{Y} \) is defined as:

\begin{equation}
\mathbf{Y} = R_\phi(\mathbf{X}, \mathbf{Z}) = \{r_\phi(x_1, z_1), r_\phi(x_2, z_2), \dots, r_\phi(x_n, z_k)\}
\end{equation}
where \( r_\phi(x_i, z_j) \) denotes the relevance score of item \( z_j \) for the input \( x_i \) as computed by the function \( R_\phi(\cdot) \). This function typically utilizes similarity measures, and retrieval methods, with the potential fine-tuning of the LLM's embeddings to improve the model's understanding of user-item interactions.

\paragraph{Non-Generative Modeling Process.} Formally, the Non-Generative approach can be expressed as:

\begin{equation}
\mathbf{Y} = R_\phi(\mathbf{X}, \mathbf{Z}) = \text{argmax}_{z \in \mathbf{Z}} \, P(z \mid \mathbf{X})
\end{equation}
where \( P(z \mid \mathbf{X}) \) represents the probability or relevance score of recommending item \( z \) given the input data \( \mathbf{X} \). The objective is to select the item \( z \) that maximizes this probability for each input \( x_i \), thus identifying the most suitable recommendation from the set \( \mathbf{Z} \).

% ============================== Non-Generative START ============================== % 
\subsubsection{LLM Retraining}

In this section, we examine three key methods under LLM Retraining: Naive Fine-Tuning, Instruction Tuning, and Low-Rank Adaptation (LoRA). Each method represents a different approach to adjusting a model’s parameters to improve its performance in recommendation tasks.

\paragraph{Naive Fine-Tuning}
\cite{2024_arXiv_Lite-LLM4Rec_Rethinking-Large-Language-Model} finds that after training, further fine-tuning the context-aware embeddings and the recommendation LLM would result in better performance. 
\cite{2024_AAAI_LLMRec_Benchmarking-Large-Language-Models} investigates the effectiveness of supervised finetuning to improve LLMs’ instruction compliance ability. The benchmark results indicates that LLMs display only moderate proficiency in accuracy-based tasks such as sequential and direct recommendation. 
\cite{2024_arXiv_NoName_improving} finds that fine-tuning an LLM for recommendation tasks enables it to learn not only the tasks but also concepts of a domain to some extent. It also shows that fine-tuning OpenAI GPT led to considerably better performance than fine-tuning Google PaLM 2 \cite{2023_Google_PaLM-2}. 
\cite{2023_CIKM_SR-Multi-Domain-Foundation_An-Unified-Search} uses LLM to extract domain-invariant features in a manner that can help deal with the cold start problems in recommendation. 
\cite{2023_arXiv_Recommender-AI-Agent} fine-tunes a 7-billion-parameter model by designing an imitation dataset derived from GPT-4, which can improve the ability of interactive recommendations. 
\cite{2023_RecSys_Leveraging-Large-Language-Models} fine-tunes an LLM with dataset-specific information in the form of prompt-completion pairs and asks the model to produce next item recommendations for test prompts.
\cite{2023_arXiv_LLMCRS_A-Large-Language-Model-Enhanced} proposes to fine-tune LLM with reinforcement learning from conversational recommender systems performance feedback for improving the performance of recommendation. 
\cite{2023_arXiv_Exploring-Fine-tuning-ChatGPT} explores fine-tuning ChatGPT by formulating the news recommendation as direct ranking and rating tasks. 
\cite{2024_NAACL_noname_aligin-large-language-modesl-with-recommendation-knowledge} proposes to align LLMs with the recommendation domain by fine-tuning with data samples that encode recommendation knowledge and also proposed auxiliary-task data samples that encode item correlations contained in users’ preferences. In the e-commerce domain, 
\cite{2023_EMNLP_LLMCRS_conver} investigates the effectiveness of combining LLM and conversational recommendation systems and fine-tunes large language models including ChatGLM and Chinese-Alpaca-7B using pre-sales dialogues. 
\cite{2024_arXiv_DEALRec_data-eff} proposes a novel data pruning method to efficiently identify the influential samples for LLM-based recommender fine-tuning, which unlocked the remarkable potential of applying LLM-based recommender models to real-world platforms. 
\cite{2024_arXiv_GACLLM_large-language} conducts supervised fine-tuning (SFT) for the LLM to activate its power in the task-related domain. This involved training the LLMs with descriptions from matched user-item pairs, allowing the LLMs to learn to align descriptions of users and items. 
\cite{2023_arXiv_BIGRec_a_bi-step} grounds LLMs in the recommendation space by fine-tuning them to generate meaningful tokens for items and subsequently identifying appropriate actual items that corresponded to the generated tokens, thereby improving the performance of recommendation systems. 
\cite{2024_arXiv_SKarREC_Learning-Structure-and-Knowledge} adopts cross-entropy loss to fine-tune its structure and knowledge-aware representation learning framework in an end-to-end manner based on concept recommendation tasks. 
\cite{2024_arXiv_LoRec_lager_lan} leverages the open-world knowledge of LLMs via supervised fine-tuning to detect fraudsters in recommender systems, enhancing robustness against poisoning attacks.
\cite{2024_arXiv_NoName_to-recommend} also fine-tunes the LLM by training it on a newly constructed dataset specifically for recommendability identification. This process involved adjusting the model’s parameters to better understand and predict whether recommendations are necessary in a given conversational context. 
\cite{2024_arXiv_LLM4POI_large} fine-tunes LLMs on a location-based social network dataset to exploit commonsense knowledge for the next point-of-interest recommendation task.
\cite{2023_arXiv_Leveraging-Large-Language-Models-in-Conversational-Recommender} focuses on tuning LLMs within their own system using large amounts of synthetically generated data.
\cite{2024_AAAI_LLM2ER-EQR_Fine-Tuning-Large} focuses on explainable recommendation task and proposes a novel LLM-based ER model denoted as LLM2ER to serve as a backbone and devise two innovative explainable quality reward models for fine-tuning such a backbone in a reinforcement learning paradigm.
\cite{2024_arXiv_Enhancing-Recommendation-Diversity-by-Re-ranking-with-Large} demonstrates how LLMs can be applied for diversity re-ranking. This work employs and compares two state-of-the-art LLM families, namely ChatGPT and Llama 2-Chat \cite{2023_LLaMA2_Open-foundation-and-fine-tuned-chat-models}, which have been fine-tuned from their respective foundation models, GPT and LLaMA, through SFT and RLHF for instruction following.
\cite{2024_WWW_NoteLLM_a-re} utilizes a Note Compression Prompt for note recommendation, compressing a note into a single special token and further learning the embeddings of potentially related notes through a contrastive learning approach.
\cite{2024_WWW_LLM-TRSR_Harnessing} constructs a prompt text that includes a user preference summary, recent user interactions, and candidate item information for an LLM-based recommender. This system was subsequently fine-tuned using supervised fine-tuning techniques to produce the final recommendation model. 
\cite{2024_arXiv_LLM4DSR} proposes LLM4DSR, a specialized approach for denoising sequential recommendations using large language models. It introduces a self-supervised fine-tuning task designed to enhance the ability of LLMs to detect noisy items within a sequence and suggest appropriate replacements. 
% Additionally, an uncertainty estimation module is developed to ensure that only high-confidence outputs are used for sequence corrections.
\cite{2024_arXiv_MixRec_Beyond-Inter-Item-Relations} proposes MixRec to enhance LLM-based sequential recommendation. Built on top of coarse-grained adaptation, MixRec is further refined with techniques such as context masking, collaborative knowledge injection, and a dynamic mixture of experts (DAMoE), enabling it to effectively manage sequential recommendation tasks.
\cite{2024_arXiv_LLMSeqPrompt_improving} designs and explores three orthogonal methods and two hybrid approaches for leveraging LLMs in sequential recommendation. Specifically, it delves into the technical aspects of each method, evaluates potential alternatives, fine-tunes them thoroughly, and assesses their overall impact.

\paragraph{Instruction Tuning}
\cite{2023_arXiv_TransRec_a_multi} proposes multi-facet identifiers consisting of numeric ID, item title, and item attribute. Based on these identifiers, and it constructs instruction data in the language space for LLMs to fine-tune, effectively combining the ranking scores from different facets. 
\cite{2023_RecSys_HKFR_hetero} constructs an instruction dataset based on the recommendation task and heterogeneous knowledge, including input, instruction, and output for personalized recommendation, it performs instruction tuning on LLM for personalized recommendations by integrating heterogeneous knowledge and recommendation tasks.
\cite{2024_AAAI_GLRec_exploring} constructs an instruction dataset to bridge the gap between pre-trained knowledge and the actual recruitment domain in online job recommendations. 
\cite{2024_WWW_IFairLRS_item_side} first represents user-item interaction data in natural language and employs instruction tuning to fine-tune LLMs, enhancing the item-side fairness of an LLM-based recommendation system. 
\cite{2024_arXiv_Llama4Rec_integrating} performs data augmentation for conventional recommendation models by leveraging instruction-tuned LLMs to alleviate data sparsity and the long-tail problem. 
\cite{2023_arXiv_RecRanker_instruction} employs adaptive user sampling to select high-quality users, facilitating the construction of the instruction-tuning dataset. This dataset is then used to train instruction-tuned LLMs for diverse ranking tasks in top-k recommendations. 
\cite{2024_WWW_E4SRec_an_elegant} designs an instruction tuning process for LLMs to stimulate their capability to follow instructions, thereby improving their ability in sequential recommendation. 
In \cite{2023_arXiv_LLaRA_Large-Language-Recommendation-Assistant}, sequential recommendation data is transformed into the instruction-tuning format for LLM on recommendation corpus. 
\cite{2023_arXiv_LLMXRec_Unlocking-the-potential} advocates for a two-stage framework called LLMXRec that decouples item recommendation from explanation generation. This study utilizes instruction tuning to enhance the precision and control of LLM-generated explanations.  
\cite{2023_arXiv_LlamaRec_LlamaRec:-Two-Stage-Recommendation-using-Large} focuses on sequential recommendation and proposes a novel LlamaRec framework for LLM-based two-stage recommendation, which proves a complete solution with both retrieval and ranking. 
\cite{2024_arXiv_EXP3RT_Review-driven-Personalized-Preference-Reasoning} presents Exp3rt, a novel recommender system based on large language models, specifically designed to harness the extensive preference information found in user and item reviews. Exp3rt effectively utilizes this detailed feedback to improve the quality and personalization of recommendations.

\paragraph{Low-Rank Adaptation (LoRA)}
\cite{2024_WSDM_ONCE_Boosting-Content-based-Recommendation} investigates the effectiveness of LoRA on the performance of open-source LLMs for content-based recommendation. 
\cite{2023_arXiv_BAHE_Breaking-the-Length-Barrier} introduces behavior aggregated hierarchical encoding (BAHE) to enhance the efficiency of LLM-based click-through rate (CTR) modeling. 
Building on LoRA, \cite{2023_RecSys_TALLRec_an_effective} efficiently incorporates supplementary information while keeping the original parameters frozen by optimizing rank decomposition matrices. 
\cite{2024_WWW_LLM-TRSR_Harnessing} applies the LoRA technique for parameter-efficient fine-tuning (PEFT), addressing the text-rich sequential recommendation problem. 
Similarly, \cite{2024_WWW_E4SRec_an_elegant} introduces an additional LoRA adapter on their proposed {gate\_proj}, {down\_proj}, and {up\_proj} modules to model the personalization of the given recommendation task. 
\cite{2024_arXiv_LoID_enhancing} adopts the LoRA strategy to train a small set of parameters for each domain, which could serve as plugins that can be seamlessly added to the target domain without further re-training.
\cite{2024_arXiv_LEADER_large} utilizes LoRA fine-tuning for medication recommendation tasks, updating sets of low-rank matrices while keeping the pre-trained weights of the LLM frozen. 
\cite{2023_arXiv_LLaRA_Large-Language-Recommendation-Assistant} proposes a curriculum prompt tuning strategy to train a LoRA, which can enhance LLMs with the behavioral knowledge encapsulated in the sequential recommenders. 
To achieve efficient unlearning while preserving high recommendation performance, \cite{2024_arXiv_APA_Exact-and-Efficient-Unlearning} employs LoRA to add pairs of rank-decomposition weight matrices to existing weights of the LLM in a plug-in manner, training only the newly added weights for learning tasks. 
\cite{2024_arXiv_E2URec_Towards-Efficient-and-Effective} proposes E2URec, the first efficient and effective unlearning method for LLM-based recommendation systems. E2URec enhances unlearning efficiency by updating only a limited set of additional LoRA parameters and improves unlearning effectiveness through a teacher-student framework.
\cite{2024_arXiv_PPLR_LLM-based-Federated-Recommendation} incorporates a dynamic balance strategy, which involves designing dynamic parameter aggregation and learning speeds for each client.
\cite{2024_arXiv_Aligning-Large-Language-Models-for-Controllable-Recommendations} introduces a novel alignment methodology for LLMs in recommender systems, significantly enhancing their ability to follow user instructions while minimizing formatting errors. 
\cite{2024_arXiv_RecLoRA} proposes RecLoRA, a model that includes a personalized LoRA module to maintain independent LoRAs for different users and a Long-Short Modality Retriever to retrieve varying history lengths for different modalities. 
\cite{2024_arXiv_MLLM-MSR_Harnessing-Multimodal-Large} proposes the Multimodal Large Language Model-enhanced Multimodal Sequential Recommendation (MLLM-MSR) model, designed to leverage large language models for improving multimodal sequential recommendations. 
\cite{2024_arXiv_GANPrompt} proposes GANPrompt, a multi-dimensional large language model prompt diversity framework based on Generative Adversarial Networks (GANs). The framework enhances the model’s adaptability and stability to diverse prompts by integrating GAN generation techniques with the deep semantic understanding capabilities of LLMs. 
\cite{2024_arXiv_DELRec_Distilling-Sequential-Pattern} proposes DELRec, a novel framework designed to extract knowledge from sequential recommendation models and enable large language models to easily comprehend and utilize this supplementary information for more effective sequential recommendations. It uses AdaLoRA to fine-tune LLMs.
\cite{2023_arXiv_CoLLM_intergrating-co} employs a two-step tuning procedure: first, fine-tuning LLM in the LoRA manner using language information exclusively to learn the recommendation task, and then specially tunes the mapping module to make the mapped collaborative information understandable and usable for LLM’s recommendation by considering the information when fitting recommendation data.

\subsubsection{LLM Reusing}

In this section, we explore three primary methods for LLM Reusing: Direct Utilizing, Prompt Tuning, and In-Context Learning. These methods involve minimal or no changes to the model's parameters, focusing instead on leveraging the pre-trained capabilities of the LLM to enhance recommender systems.

\paragraph{Direct Utilizing}

\cite{2023_arXiv_DOKE-Knowledge-Plugins:Enhancing-Large} proposes a general paradigm that augments LLMs with Domain-specific knowledge to enhance their performance on practical applications, namely DOKE. This paradigm relies on a domain knowledge extractor, which involves preparing effective knowledge for the task, selecting the knowledge for each specific sample, and expressing the knowledge in a way that is understandable by LLMs.
\cite{2023_arXiv_DRDT_DRDT:Dynamic-Reflection-with-Divergent} introduces the Dynamic Reflection with Divergent Thinking (DRDT) in a retriever-reranker framework, a novel approach that effectively integrates collaborative signals and the temporal evolution of user preferences into sequential recommendation tasks using LLMs. 
\cite{2023_arXiv_NIR_Zero-Shot-Next-item-Recommendation-using-Large-Pretrained} proposes a prompting strategy called Zero-Shot Next-Item Recommendation (NIR) prompting that directs LLMs to make next-item recommendations. Specifically, the NIR-based strategy involves using an external module to generate candidate items based on user filtering or item filtering. 
\cite{2023_ICDM-Workshop_Zero-shot-Recommendation} presents a method for zero-shot recommendation of multimodal non-stationary content that leverages recent advancements in the field of generative AI.
\cite{2023_RecSys_Large-Language-Models-are-Competitive-Near-Cold-start} proposes various prompting methods for LLMs for the task of language-based item recommendation.
\cite{2023_RecSys_MINT_Large-Language-Model-Augmented-Narrative-Driven} presents Mint, a data augmentation method for the narrative-driven recommendation (NDR) task. Mint re-purposes historical user-item interaction datasets for NDR by using a 175B parameter large language model to author long-form narrative queries while conditioning on the text of items liked by users.
\cite{2024_arXiv_ChatGPT-for-Conversational-Recommendation:Refining} investigates the effectiveness of ChatGPT as a top-n conversational recommendation system. A comprehensive pipeline is built around ChatGPT to simulate realistic user interactions in probing the model for recommendations.
\cite{2024_arXiv_GPT-FedRec_Federated-Recommendation} proposes GPTFedRec, a federated recommendation framework leveraging ChatGPT and a novel hybrid Retrieval Augmented Generation (RAG) mechanism. 
\cite{2024_arXiv_Re2LLM_Re2LLM:Reflective-Reinforcement-Large-Language-Model-for-Session-based} introduces a novel learning paradigm that goes beyond in-context learning and fine-tuning of LLMs, effectively bridging general LLMs with specific recommendation tasks. 
\cite{2024_NAACL_RecMind_Large-Language-Model-Powered} introduces RecMind, a LLM-powered agent designed for general recommendation purposes. RecMind operates without the need for fine-tuning to adapt to different domains, datasets, or tasks. RecMind incorporates a novel self-inspiring (SI) planning technique.
\cite{2024_arXiv_RTA_Reindex-Then-Adapt} presents a Reindex-Then-Adapt (RTA) framework that converts multi-token item titles into single tokens within LLMs and subsequently adjusts the probability distributions for these single-token item titles. 
\cite{2024_arXiv_Tempura_Improve-Temporal-Awareness-of-LLMs} proposes three prompting strategies to utilize temporal information within historical interactions for LLM-based sequential recommendation. This study incorporates explicit structure analysis in input sequences as additional prompts, specifically temporal cluster analysis.
\cite{2024_arXiv_UniLLMRec_Tired-of-Plugins?} introduces an LLM-based end-to-end recommendation framework called UniLLMRec. UniLLMRec integrates multi-stage tasks such as recall, ranking, and re-ranking through a chain-of-recommendations approach. 
\cite{2024_arXiv_LLMmove_Where-to-Move-Next:Zero-shot} focuses on harnessing the capabilities of Large Language Models for the zero-shot next POI recommendation task. The approach considers users' long-term and current preferences, geographic spatial distance, and sequential transitions in user mobility behaviors. 
\cite{2024_ECIR_Large-Language-Models-are-Zero-Shot-Rankers}  formalizes the recommendation problem as a conditional ranking task, using sequential interaction histories as conditions and items retrieved by other models as candidates. 
\cite{2024_PAKDD_InteraRec} presents InteraRec, an innovative screenshot-based user recommendation system. InteraRec captures real-time, high-frequency screenshots of web pages as users browse. Utilizing the capabilities of MLLMs, it analyzes these screenshots to derive meaningful insights into user behavior and uses relevant optimization tools to provide personalized recommendations.
\cite{2024_SIGIR_BiLLP_Large-Language-Models-are-Learnable-Planners} proposes leveraging the exceptional planning capabilities of Large Language Models to handle sparse data for long-term recommendations. This work introduces a bilevel learnable LLM planner framework, which employs a set of LLM instances. 
\cite{2024_WSDM_Logic-Scaffolding_Personalized-aspect-instructed} proposes a framework called Logic-Scaffolding, which combines the concepts of aspect-based explanation and chain-of-thought prompting to generate explanations through intermediate reasoning steps.
\cite{2024_WWW_ReLLa_ReLLa:Retrieval-enhanced-Large-Language} performs semantic user behavior retrieval (SUBR) to enhance the data quality of testing samples for zero-shot recommendation, significantly reducing the difficulty for LLMs to extract essential knowledge from user behavior sequences.
\cite{2024_arXiv_RDRec_Rationale-Distillation} proposes a compact RDRec model to learn the underlying rationales for interactions generated by an LLM. By learning rationales from all related reviews, RDRec effectively specified user and item profiles by designing a prompt template for recommendations. 
\cite{2024_arXiv_LLMHG_Llm-guided-multiview-hypergrap-learning} facilitates nuanced LLM-based user profiling while still accounting for sequential user behavior. By generating and refining prompts that guide the hypergraph learning process.
\cite{2024_arXiv_LLM_InS_large} proposes an LLM Interaction Simulator to model users’ behavior patterns based on content aspects. This prompt-based simulator allowed recommender systems to simulate vivid interactions for each cold item and transform them from cold to warm items directly. 
\cite{2024_WSDM_LLMRec} enables LLMs to generate user and item attributes that were not originally part of the dataset by using prompts derived from the dataset’s interactions and side information. 
\cite{2023_arXiv_LKPNR_LKPNR:LLM-and-KG-for-Personalized} merges a Large Language Model with a Knowledge graph (KG). By incorporating the general news encoder, the LLM's strong contextual understanding enables the generation of news representations rich in semantic information.
\cite{2023_arXiv_LLM4Vis_Explainable-visualization-recommendation} introduces LLM4Vis, an innovative ChatGPT-based prompting method that provides visualization recommendations and produces human-like explanations with the use of only a few demonstration examples.
\cite{2024_AAAI_LLMRG_Improving-Recommendations-through-LargeLanguageModel} introduces LLMRG, which employs LLMs to build personalized reasoning graphs. This approach illustrates how LLMs can enhance logical reasoning and interpretability in recommendation systems without the need for additional information.
\cite{2024_SIGIR_PO4ISR_Large-Language-Models-for-Intent-Driven-Session-Recommendation}  introduces a straightforward yet powerful paradigm, PO4ISR, that leverages the capabilities of LLMs to enhance ISR through prompt optimization. 
\cite{2023_arXiv_SAGCN_Understanding-Before-Recommendation} presents a chain-based prompting method, leveraging the deep semantic comprehension of large language models to reveal semantic aspect-aware interactions. This approach offers more detailed insights into user behaviors at a fine-grained semantic level.
\cite{2023_arXiv_TF-DCon_Levaraging-Large-Language-Models(LLMS)-to-Empower-Training-Free-Dataset-Condensation} proposes TF-DCon framework and takes inspiration from the exceptional text comprehension and generation capabilities of large language models and utilizes them to enhance the generation of textual content during condensation.
\cite{2024_arXiv_CoRAL_CoRAL:-Collaborative-Retrieval-Augmented-Large-Language} presents CoRAL, a method designed to enhance long-tail recommendations within traditional collaborative filtering-based systems. It effectively addresses the challenges posed by data sparsity and imbalance, which often limit the performance of collaborative filtering methods. 
\cite{2024_arXiv_CSRec_Common-Sense-Enhanced-Knowledge-based-Recommendation} presents a novel framework, CSRec, which develops an LLM-based common sense knowledge graph and incorporates it into the recommendation system using a mutual information maximization (MIM)-based knowledge fusion technique.
\cite{2024_arXiv_DynLLM_When-Large-Language-Models} introduces a novel task, LLM-augmented dynamic recommendation using continuous time dynamic graphs, and proposes the DynLLM model to effectively integrate LLM-augmented data with temporal graph information. 
\cite{2024_SIGIR_FineRec_FineRec:Exploring-Fine-grained-Sequential-Recommendation}  introduces a novel framework, FineRec, designed to explore fine-grained sequential recommendations by mining attribute-opinions from reviews. 
\cite{2024_arXiv_LLM4SBR_LLM4SBR:A-Lightweight-and-Effective-Framework-for-Integrating} presents a scalable two-stage LLM enhancement framework (LLM4SBR) specifically designed for session-based recommendation (SBR). examines the feasibility of integrating LLM with SBR models, focusing on both effectiveness and efficiency. In the context of short sequence data, LLM can infer preferences directly through its language understanding capability, even without fine-tuning.
\cite{2024_arXiv_LLM-KERec_Breaking-the-Barrier-Utilizing} focuses on industrial recommendation systems and proposes LLM-KERec, it employs a large language model to determine the existence of complementary relationships between two entities and constructs a complementary graph.
\cite{2024_SIGIR_LRD_Sequential-Recommendation} proposes a novel relation-aware sequential recommendation framework with Latent Relation Discovery (LRD). Different from previous relation-aware models that rely on predefined rules, it proposes to leverage the Large Language Model to provide new types of relations and connections between items.
\cite{2024_arXiv_News-Recommendation-with-Category-Description-by} presents a novel method that automatically generates informative category descriptions using a large language model without requiring manual effort or domain-specific knowledge, and integrates them into recommendation models as supplementary information.
\cite{2024_arXiv_PAP-REC_Personalized-Automatic-Prompt} introduces PAP-REC, a framework designed to generate personalized automatic prompts for recommendation language models, addressing the inefficiency and ineffectiveness of manually crafted prompts. 
\cite{2024_WWW_Large-Language-Models-as-Data-Augmenters-for-Cold-Start-Item-Recommendation} proposes leveraging LLMs as data augmenters to address the knowledge gap associated with cold-start items during training. By utilizing LLMs, it infers user preferences for cold-start items based on the textual descriptions of users' historical behaviors and the descriptions of new items.
\cite{2024_arXiv_MMREC-LLM-Based-Multi-Modal-Recommender} investigates the potential of LLMs to enhance the understanding and utilization of natural language data within recommendation contexts.
\cite{2024_arXiv_DaRec_A-Disentangled-Alignment} proposes DaRec, a novel plug-and-play disentangled alignment framework for integrating recommendation models with large language models.
\cite{2024_arXiv_X-Reflect_Cross-Reflection-Prompting-for-Multimodal-Recommendation} introduces X-REFLECT, a novel CrossReflection Prompting framework. This method prompts large multimodal models (LMMs) to process textual and visual information concurrently, explicitly identifying and reconciling any supportive or conflicting elements between these modalities.
\cite{2024_arXiv_LLM4MSR} proposes LLM4MSR, an efficient and interpretable LLM-enhanced paradigm. It uses LLMs to extract multi-level knowledge, such as scenario correlations and cross-scenario user interests, via a tailored prompt without fine-tuning. 
% Hierarchical meta networks are then used to generate multi-level meta layers, enhancing scenario-aware and personalized recommendations.
\cite{2023_ICML_NoName_LLMbased} aims to measure the effectiveness of using item aspects-justifications for users' purchase intentions-generated by LLMs to improve ranking task results. To achieve this, prompts are carefully designed to derive aspects for items from their textual data in an ecommerce setting.
\cite{2023_arXiv_KAR_towards} introduces factorization prompting to elicit accurate reasoning on user preferences. The generated reasoning and factual knowledge were effectively transformed and condensed into augmented vectors by a hybrid-expert adaptor to be compatible with the recommendation task.
\cite{2024_arXiv_LLM-Rec_Personalize} introduces LLM-Rec, a novel approach that incorporates four distinct prompting strategies: basic prompting, recommendation-driven prompting, engagement-guided prompting, and a combination of recommendation-driven and engagement-guided prompting.

\paragraph{Prompt Tuning}
\cite{2024_WWW_PMG_Personalized-Multimodal-Generation} focuses on the personalized multimodal generation problem and proposes PMG. It incorporates multimodal tokens as learnable parameters into the embedding table and then utilizes a linear layer to align the embedding space of the LLM with that of the generator.
\cite{2024_arXiv_LFM_language-based-user} tests Llama 2-7B, Llama 2-13B, and Sakura-SOLAR 10.7B\footnote{\url{https://huggingface.co/kyujinpy/Sakura-SOLAR-Instruct}} in instruct (chat) mode, using zero-shot prompt tuning. Hard prompt tuning was performed using a validation dataset to enhance model accuracy, runtime, and reliability.
\cite{2023_arXiv_Prompt-Tuning-Large-Language-Models-on-Personalized-Aspected-Extraction} proposes an end-to-end framework that combines the learning of LLM-based personalized aspect extraction through prompt tuning with aspect-based recommendation, resulting in more effective recommendations.
\cite{2023_arXiv_RecPromp_A-Prompt-Tuning} introduces a prompt tuning framework for news recommendation using LLMs, and it distinctively integrates a prompt optimizer with an iterative bootstrapping method to improve LLM-based recommendation strategies.

\paragraph{In-Context Learning (ICL)}
\cite{2023_arXiv_Chat-rec_Towards-interactive-and-explainable} introduces a novel paradigm called Chat-Rec, which innovatively enhances LLMs for developing conversational recommender systems by transforming user profiles and historical interactions into prompts. Chat-Rec has been shown to effectively learn user preferences and establish connections between users and products through in-context learning, thereby making the recommendation process more interactive and explainable.
\cite{2024_arXiv_DRE_Generating-Recommendation-Explanations} introduces Data-level Recommendation Explanation (DRE), a non-intrusive framework designed to provide explanations for black-box recommendation models. This work proposes leveraging the in-context learning and reasoning abilities of LLMs to align the explanation module with the recommendation module. It employs an in-context learning approach and instructs the LLM to generate a logically coherent recommendation explanation that aligns with the recommendation system and corresponds to user attention preferences.
\cite{2024_arXiv_Efficient-and-Responsible-Adaptation-of-Large-Language} proposes a hybrid task allocation framework that utilizes the capabilities of both LLMs and traditional recommendation systems. By adopting a two-phase approach to improve robustness to sub-populations, the framework promotes a strategic assignment of tasks for efficient and responsible adaptation of LLMs.
\cite{2023_RecSys_Uncovering-ChatGPT's-Capabilities-in-Recommender} aims to enhance ChatGPT’s recommendation capabilities by aligning it with traditional information retrieval (IR) ranking methods, such as point-wise, pair-wise, and list-wise ranking. This work adopts in-context learning and instruction learning, and it expresses different abilities as different tasks with domain-specific prompts.
\cite{2024_SSRN_New-Community-Cold-Start-Recommendation} tackles the New Community Cold-Start (NCCS) problem by proposing a novel recommendation method that leverages the extensive knowledge and powerful inference capabilities of Large Language Models. It selects In-Context Learning (ICL) as the prompting strategy and designs a coarse-to-fine framework to efficiently choose demonstration examples for creating effective ICL prompts. 
\cite{2024_arXiv_LANE_Logic-Alignment} proposes LANE, an effective strategy that aligns large language models with online recommendation systems without requiring additional tuning of the LLMs. This approach reduces costs while enhancing the explainability of recommendations. This work leverages the exceptional in-context learning capabilities of large language models and carefully designs a zero-shot prompt template to extract users' multiple preferences.

% ============================== Non-Generative END ============================== % 

% ============================== Generative START ============================== %  

% ---------------- tables: Generative Models --------------
\begin{table*}[t!]
    \centering
    \caption{The representative LLM-based works for generative recommendations.}
    \resizebox{0.98\textwidth}{!}{
        \begin{tabular}{c|c|c|c|c}
        % \hline
        \toprule
            \textbf{Model/Paper} & \textbf{Task/Domain} & \textbf{Data Modality} & \textbf{Main Techniques} & \textbf{Source Code} \\
            \midrule
            
            \citet{2023_arXiv_Multiple-KV} 
            & sequential recommendation 
            & multiple key-value data 
            & pre-train, instruction tuning 
            & ~ \\ \hline
            
            \citet{2023_CIKM_LLM-as-Zero-Shot-Conversational-Recommendation} 
            & \makecell{zero-shot conversational recommendation} 
            & \makecell{text\\(conversational recommendation\\dataset)} 
            & prompt 
            & \parbox{6cm}{\href{https://github.com/AaronHeee/LLMs-as-Zero-Shot-Conversational-RecSys}{https://github.com/AaronHeee/LLMs-as-Zero-Shot-Conversational-RecSys}} \\ \hline
            
            \citet{2023_Electronics_Bookgpt_A-general-framework} 
            & book recommendation 
            & interaction, text 
            & prompt 
            & \url{https://github.com/zhiyulee-RUC/bookgpt} \\ \hline
            
            \citet{2023_SIGIR-AP_How-to-Index-Item-IDs} 
            & sequential recommendation 
            & interaction, text 
            & item ID indexing 
            & \url{https://github.com/Wenyueh/LLM-RecSys-ID} \\ \hline
            
            \citet{2024_arXiv_Supporting-student-decisions} 
            & learning recommendation 
            & graph data, text 
            & ~ 
            & ~ \\ \hline
            
            \citet{2023_arXiv_GPT4Rec} 
            & next-item prediction 
            & interaction, item title 
            & prompt, pre-train, fine-tune 
            & ~ \\ \hline
            
            \citet{2024_arXiv_Leveraging-Edge-Information} 
            & item recommendation 
            & interaction 
            & prompt, pre-train, fine-tune 
            & \url{https://github.com/anord-wang/LLM4REC.git} \\ \hline
            
            \citet{2024_arXiv_RecGPT_Sequential-Recommendation} 
            & sequential recommendation 
            & sequences of words 
            & prompt, pre-train, fine-tune 
            & ~ \\ \hline
            
            \citet{2024_arXiv_RecGPT_Text-based-Recommendation} 
            & \makecell{rating prediction, \\sequential recommendation} 
            & text 
            & pre-train, fine-tune 
            & \url{https://github.com/VinAIResearch/RecGPT} \\ \hline
            
            \citet{2024_ECIR_GenRec} 
            & movie recommendation 
            & interaction, textual-information 
            & prompt, pre-train, fine-tune 
            & \url{https://github.com/rutgerswiselab/GenRec} \\ \hline
            
            \citet{2024_SIGIR_IDGenRec} 
            & \makecell{sequential recommendation,\\zero-shot recommendation} 
            & interaction, text 
            & natural language generation 
            & \url{https://github.com/agiresearch/IDGenRec} \\ \hline
            
            \citet{2024_WWW_CLLM4Rec} 
            & item recommendation 
            & interaction, text 
            & prompt, pre-train, fine-tune 
            & \url{https://github.com/yaochenzhu/llm4rec} \\ \hline
            
            \citet{2024_WWW_PMG_Personalized-Multimodal-Generation} 
            & personalized multimodal generation 
            & text, image, audio, etc 
            &\makecell{prompt, pre-train,\\ Prompt Tuning (P-Tuning V2 \citet{2021_arXiv_P-Tuning_V2})} 
            & \parbox{6cm}{\href{https://github.com/mindspore-lab/models/tree/master/research/huawei-noah/PMG}{https://github.com/mindspore-lab/models/tree/master/research/huawei-noah/PMG}} \\ \hline
            
            \citet{2024_arXiv_CALRec} 
            & sequential recommendation 
            & interaction, text 
            & pre-train, fine-tune, contrastive learning
            & ~ \\ \hline
            
            \citet{2024_WSDM_ONCE_DIRE-GENRE} 
            & \makecell{content-based recommendation\\(news recommendation,\\book recommendation)} 
            & interaction, text 
            & prompt 
            & \url{https://github.com/Jyonn/ONCE} \\
            \bottomrule

            % ~ & ~ & ~ & ~ & ~ \\ \hline
        \end{tabular}
        \label{tab: generative llm-based models}
    }
\end{table*}

\subsection{Generative LLM-based Approaches}

The Generative approach involves using LLMs to generate new content or recommendations based on the input data. This method is particularly useful in scenarios where novel user interactions or items are required. Formally, let $\mathbf{X} = \{x_1, x_2, \dots, x_n\}$ represent the input data, and let $\mathbf{Y} = \{y_1, y_2, \dots, y_m\}$ represent the generated recommendations. The goal of the Generative approach is to learn a mapping function $G_\theta(\cdot)$ parameterized by $\theta$, such that:

\begin{equation}
\mathbf{Y} = G_\theta(\mathbf{X}) = \{g_\theta(x_1), g_\theta(x_2), \dots, g_\theta(x_n)\}
\end{equation}
where $G_\theta(\cdot)$ leverages the internal knowledge and structure of the LLM to produce recommendations that are not merely a reflection of the input data but are instead novel and contextually relevant.

\textbf{Generative Modeling Process.} Given an input sequence $\mathbf{X}$, the Generative approach can be described as follows:

\begin{equation}
\mathbf{Y} = G_\theta(\mathbf{X}) = P(y_1, y_2, \dots, y_m \mid x_1, x_2, \dots, x_n)
\end{equation}

Here, $P(y_1, y_2, \dots, y_m \mid x_1, x_2, \dots, x_n)$ represents the probability distribution over the generated recommendations, conditioned on the input sequence $\mathbf{X}$. The LLM generates each $y_i$ based on this learned distribution. Table \ref{tab: generative llm-based models} lists representative papers on generative LLM-based recommendations.

\subsubsection{LLM Retraining}

In this section, we examine three key methods under LLM Retraining: Naive Fine-Tuning, Instruction Tuning, and LoRA.

\paragraph{Naive Fine-Tuning}

\cite{2023_arXiv_GPT4Rec} introduces GPT4Rec, a new and versatile generative framework. Initially, it generates hypothetical "search queries" based on the item titles in a user's history, and then it retrieves items for recommendation by searching these queries. To effectively capture user interests across various aspects and levels of detail, enhancing both relevance and diversity, the framework employs a multi-query generation technique using beam search. It fine-tunes the selected GPT-2 \cite{2019_OpenAI-blog_GPT-2} model, which has 117 million parameters, features a sophisticated transformer architecture, and is pre-trained on a vast language corpus. This process enables us to capture user interests and item content information effectively.
\cite{2024_arXiv_Leveraging-Edge-Information} proposes a new prompt mechanism that can transform the relationship between users and items, as well as the background information of items, into natural language form, and introduces a new fusion method to effectively exploit the connection information in graph (i.e, the edge information in graph structure). It pre-trains graph attentive LLM (graph attentive GPT-2 \cite{2019_OpenAI-blog_GPT-2} model) with crowd contextual prompts and fine-tunes graph attentive LLM with personalized predictive prompts. 
\cite{2024_arXiv_RecGPT_Sequential-Recommendation} focuses on sequential recommendation and proposes modeling user behavior sequences with a personalized prompt via the ChatGPT training paradigm, it proposes a new framework called RecGPT. It pre-trains a personalized auto-regressive generative model by introducing user IDs modular,and then fine-tunes the pre-trained model by introducing segment IDs to generate personalized prompts. 
\cite{2024_arXiv_RecGPT_Text-based-Recommendation} focuses on text-based recommendation, and introduces domain-adapted and fully-trained large language model named RecGPT-7B, and its instruction-following variant, RecGPT-7B-Instruct. it pre-trains RecGPT-7B using a relatively large recommendation-specific corpus of 20.5B tokens, while RecGPT-7B-Instruct is the model output by further fine-tuning RecGPT-7B on a dataset of 100K+ instructional prompts and their responses.
\cite{2024_SIGIR_IDGenRec} proposes IDGenRec and selects a T5 \cite{2020_JMLR_T5} model originally trained for article tag generation and fine-tunes it with recommendation objectives. It focuses on standard sequential recommendation tasks and zero-shot recommendation scenarios.
\cite{2024_WWW_CLLM4Rec} introduces CLLM4Rec, a generative recommendation system that tightly integrates the ID paradigm with the LLM paradigm. It presents an innovative soft+hard prompting strategy to effectively pre-train CLLM4Rec on heterogeneous tokens representing historical interactions and user/item features. Additionally, it proposes a recommendation-oriented fine-tuning strategy to predict hold-out items.
\cite{2024_arXiv_CALRec} introduces CALRec, a novel contrastive-learning-assisted two-stage training framework tailored for sequential recommendation tasks utilizing the PaLM-2 large language model as the backbone. This framework incorporates meticulously crafted templates inspired by few-shot learning principles and a unique quasi-round-robin BM25 retrieval strategy.

\paragraph{Instruction Tuning}
\cite{2023_arXiv_Multiple-KV} aims to implement sequential recommendation based on multiple key-value data by incorporating recommendation systems with LLM. In particular, it instructs tuning a prevalent open-source LLM (LLaMA 7B \cite{2023_arXiv_LLaMA}) in order to inject domain knowledge of RS into the pre-trained LLM. Given that this work uses multiple key-value strategies, it becomes challenging for the LLM to effectively learn from these keys. Therefore, it has devised novel shuffle and mask strategies as an innovative approach to data augmentation.

\paragraph{LoRA (Low-Rank Adaptation)}
\cite{2024_ECIR_GenRec} highlights the promising paradigm of generative recommendation and proposes GenRec model which incorporates textual information to enhance the generative recommendation performance. It selects the LLaMA language model as the backbone. LLaMA model is pre-trained on an expansive language corpus, providing a valuable resource to efficiently capture both user interests and item content information. To conserve GPU memory, it adopts the LLaMA-LoRA architecture for fine-tuning and inference tasks.

\subsubsection{LLM Reusing}

In this section, we explore two primary methods for LLM Reusing: Direct Utilizing and Prompt Tuning.

\paragraph{Direct Utilizing}
\cite{2023_SIGIR-AP_How-to-Index-Item-IDs} explores various ID creation and indexing methods, examining three basic indexing approaches: Random Indexing, Title Indexing, and Independent Indexing, while highlighting their limitations. The study underscores the significance of choosing an appropriate indexing method for foundation recommendation models, as it significantly influences model performance. Additionally, four straightforward yet effective indexing methods are investigated: Sequential Indexing, Collaborative Indexing, Semantic Indexing, and Hybrid Indexing.
\cite{2024_arXiv_Supporting-student-decisions} proposes an approach to employ chatbots as conversation mediators and sources of controlled, limited generation of explanations. This method aims to harness the capabilities of LLMs while simultaneously mitigating their potential risks. The proposed LLM-based chatbot is designed to assist students in comprehending learning-path recommendations. It utilizes a knowledge graph (KG) as a human-curated information source to regulate the LLM's output by defining the context of its prompts. Additionally, a group chat approach is implemented to link students with human mentors, either upon request or when situations exceed the chatbot's predefined capabilities.
\cite{2023_CIKM_LLM-as-Zero-Shot-Conversational-Recommendation} explores the use of Large Language Models for zero-shot conversational recommendation systems. And it introduces a simple prompting strategy to define the task description, format requirement and conversation context for an LLM. This work then post-processes the generative results into ranked item lists with processor.
\cite{2023_Electronics_Bookgpt_A-general-framework} focuses on ChatGPT as the modeling subject, integrating LLM technology into the standard scenario of understanding and recommending book resources for the first time and putting it into practice. By developing a ChatGPT-like book recommendation system (BookGPT) framework, it seeks to apply ChatGPT to recommendation modeling for three key tasks: book rating recommendation, user rating recommendation, and book summary recommendation. Additionally, it investigates the feasibility of LLM technology in the context of book recommendations. It discusses construction ideas and prompt engineering methods for three subtasks. Empirical research has been conducted to verify the feasibility of two distinct prompt modeling methods: zero-shot modeling and few-shot modeling.
\cite{2024_WSDM_ONCE_DIRE-GENRE} employs prompting techniques to enrich the training data at the token level for closed-source LLMs. This work introduces a generative recommendation method called GENRE. By developing diverse prompting strategies, it enhances the available training data and obtain more informative textual and user features, leading to better performance in subsequent recommendation tasks.

\paragraph{Prompt Tuning}
\cite{2024_WWW_PMG_Personalized-Multimodal-Generation} focuses on personalized multimodal generation problem and proposes PMG. It incorporates multimodal tokens as learnable parameters into the embedding table and then utilizes a linear layer to align the embedding space of the LLM with that of the generator. 
Additionally, it uses P-Tuning V2 \cite{2021_arXiv_P-Tuning_V2} to fine-tune the LLM specifically for the generation task, thereby enhancing its generative capabilities. During each inference, the multimodal tokens are appended to the user behavior prompt. The soft preference embeddings are then generated by passing these augmented inputs through the LLM (enhanced with P-Tuning V2) and a linear layer.

\subsubsection{Components or Strategies for Generative Recommendation}
\cite{2024_arXiv_LETTER} focuses on the item tokenization problem and comprehensively analyzes the necessary features of an ideal identifier, proposing a novel learnable tokenizer named LETTER to adaptively learn identifiers encompassing hierarchical semantics, collaborative signals, and code assignment diversity.
% ============================== Generative END ============================== %  

\section{Industrial Deploying}
\label{sec_Industrial Deploying}

Deploying LLM-based recommender systems in large-scale industrial settings involves several crucial aspects. This section examines the main focus areas for LLM-based systems, including their approach to large-scale industrial scenarios, acceleration, cold start, implementing dynamic updates, and meeting various business customization requirements. By reviewing these focus areas, we can better appreciate the current advancements and practical considerations in the deployment of LLM-based recommender systems in real-world industrial applications. Table \ref{tab: industrial models} lists representative works that could be implemented in industry.

% ---------------- tables: Industrial Models --------------
\begin{table*}[t!]
    \centering
    \caption{The representative LLM-based works designed for deployment in industry.}
    \resizebox{0.98\textwidth}{!}{
    \begin{tabular}{c|c|c|c}
    % \hline
    \toprule
        \textbf{Model/Paper} & \textbf{Company} & \textbf{Task/Domain} & \textbf{Hightlights} \\
        \midrule
        
        % 1.规模化
        LLM-KERec,~\citet{2024_arXiv_LLM-KERec_Breaking-the-Barrier-Utilizing} & Ant Group & E-commerce Recommendation& \makecell{Constructs a complementary\\knowledge graph by LLMs}
        \\ \hline
        LSVCR,~\citet{2024_arXiv_LSVCR_A-Large-Language-Model-Enhanced} & KuaiShou & Video Recommendation & \makecell{Sequential recommendation model\\and supplemental LLM recommender} 
        \\ \hline
        RecGPT,~\citet{2024_arXiv_RecGPT_Sequential-Recommendation}& KuaiShou & Sequential Recommendation & \makecell{Models user behavior sequences using\\personalized prompts with ChatGPT.} 
        \\ \hline
        % 重复，文章已有\citet{2024_arXiv_LEARN_Knowledge-Adaptation-from} 
        SAID,~\citet{2024_WWW_SAID_Enhancing-sequential-recommendation}  & Ant Group  & Sequential Recommendation & \makecell{Explicitly learns Semantically Aligned item\\ID embeddings based on texts by utilizing LLMs.}
        \\ \hline
        % 重复，文章已有\citet{2024_arXiv_HSTU_Actions-speak-louder} 
        BAHE,~\citet{2023_arXiv_BAHE_Breaking-the-Length-Barrier}& Ant Group &  CTR Prediction & \makecell{Uses LLM's shallow layers for user behavior embeddings\\and deep layers for behavior interactions.}  
        \\ \hline
        % 未应用落地 
        \citet{2024_arXiv_LLM4SBR_LLM4SBR:A-Lightweight-and-Effective-Framework-for-Integrating} & HuaWei & Session-based Recommendation & \makecell{In short sequence data, LLM can infer\\ preferences directly leveraging its language\\ understanding capability without fine-tuning}
        \\ \hline

        % Efficiency
        DARE,~\citet{2024_arXiv_DARE_A-Decoding-Acceleration-Framework-for-Industrial-Deployable} & HuaWei & CTR Prediction & \makecell{Identifies the issue of inference efficiency \\during deploying LLM-based recommendations\\ and introduce speculative decoding to accelerate \\recommendation knowledge generation.}
        \\ \hline
        
        % 冷启动
        \citet{2023_CIKM_SR-Multi-Domain-Foundation_An-Unified-Search} & Ant Group & Multi-domain Recommendation & \makecell{LLM is applied to the S\&R multi-domain foundation model\\to extract domain-invariant text features.}  \\ \hline
        LEARN,~\citet{2024_arXiv_LEARN_Knowledge-Adaptation-from} & Kuaishou & Sequential Recommendation & \makecell{Integrates the open-world knowledge\\encapsulated in LLMs into RS.}  \\ \hline
        
        % 动态更新
        \citet{2024_arXiv_DynLLM_When-Large-Language-Models} & Alibaba & E-commerce Recommendations & \makecell{Generates user profiles based on textual historical purchase\\records and obtaining users' embeddings by LLM.}
        \\ \hline
        HSTU,~\citet{2024_arXiv_HSTU_Actions-speak-louder} & Meta &  Generating user action sequence & \makecell{Explores the scaling laws of RS;\\Optimising model architecture to accelerate inference.} 
        \\ \hline
        COSMO,~\citet{2024_SIGMOD-Companion_COSMO} & Amazon & \makecell{Semantic Relevance and Session-based \\Recommendation} & \makecell{Is the first industry-scale knowledge system that\\ adopts LLM to construct high-quality knowledge\\ graphs and serve online applications}
        \\ \hline
        
        % 模型定制化
        SINGLE,~\citet{2024_WWW_SINGLE_Modeling-User-Viewing-Flow}  & Taobao, Alibaba & Article Recommendation & \makecell{Summaries long articles and the user constant preference \\from view history by gpt-3.5-turbo or ChatGLM-6B.} 
        \\ \hline
        \citet{2024_arXiv_Ad-Recommendation} & Tencent & Ad Recommendation & \makecell{Obtains user or items embeddings by LLM.} 
        \\ \hline
        NoteLLM,~\citet{2024_WWW_NoteLLM_a-re} & xiaohongshu & Item-to-item Note Recommendation & \makecell{Obtains article embeddings and generate \\hashtags/categories infomation by LLaMA-2.}
        \\ \hline
        Genre Spectrum,~\citet{2023_RecSys_Beyond-Labels-Leveraging} & tubi & Movie \& TV series Recommendation  & \makecell{Obtains content metadata embeddings by LLM.} 
        \\ \hline
        TRAWL,~\citet{2024_arXiv_TRAWL_External-Knowledge-Enhanced}  & WeChat, Tencent & Article Recommendation  & \makecell{Uses Qwen1.5-7B extract knowledge from articles.}
        \\ \hline
        HKFR,~\citet{2023_RecSys_HKFR_hetero}& Meituan & Catering Recommendation  & \makecell{Uses heterogeneous knowledge fusion for recommendations.}
        \\ 
        \bottomrule
    \end{tabular}
    \label{tab: industrial models}
    }
\end{table*}

\subsection{Large-Scale Industrial Scenarios}

In large-scale industrial applications, deploying LLM-based recommender systems brings significant complexity due to the vast data volumes and constantly changing user and item dynamics. These environments demand efficient processing of extensive feature spaces while accommodating diverse and evolving business needs. The scale of these systems amplifies the need for a balance between computational efficiency and maintaining high-quality recommendations.

Traditional recommendation pipelines are composed of smaller, more specialized models, which are easier to manage in terms of cost and updates. In contrast, the resource demands of LLMs — particularly in training and inference — make their large-scale deployment more challenging. Instead of entirely replacing traditional models, LLMs are increasingly being integrated as components to enhance overall performance. 
\cite{2024_arXiv_LLM-KERec_Breaking-the-Barrier-Utilizing} proposes LLM-KERec method combines the efficient collaborative signal processing capabilities of traditional models with LLMs and complementary graphs. This approach not only reduces the homogeneity of recommendation results from traditional models but also improves overall click-through rates and conversion rates, enabling the large-scale application of LLMs in industrial scenarios.
\cite{2024_arXiv_LSVCR_A-Large-Language-Model-Enhanced} points out that the computational costs of LLMs make them difficult to deploy effectively as online recommendation systems in large-scale industrial contexts. Therefore, this work proposes using LLMs solely during the training phase to supplement and enhance the semantic capabilities of our recommendation backbone.
To address the issue of large models being too heavy for online recommendation services, \cite{2024_arXiv_RecGPT_Sequential-Recommendation} explores how to integrate ChatGPT flexibly and effectively into RS. They abandons the natural language form, and use ChatGPT's model structure and training paradigm for project sequence prediction.
\cite{2024_arXiv_LEARN_Knowledge-Adaptation-from} states that reasoning LLMs or fine-tuning LLMs based on user interaction history is impractical in industrial scenarios. To address the computational challenges posed by processing large amounts of user interaction history, the CEG module employs a pre-trained LLM as an item encoder rather than a user preference encoder, thereby reducing computational overhead.
\cite{2024_WWW_SAID_Enhancing-sequential-recommendation} introduces SAID, a framework that leverages LLMs to explicitly learn text-based semantic alignment for item ID embeddings. This approach reduces the resources required in industrial scenarios.
\cite{2024_arXiv_HSTU_Actions-speak-louder} proposes a new recommendation system paradigm, Generative Recommenders, which redefines the recommendation problem as a sequential transduction task within a generative modeling framework. Using GRs, the complexity of the deployed model increased by 285 times, while using less inference computation.
\cite{2023_arXiv_BAHE_Breaking-the-Length-Barrier} believes that a key obstacle to deploying LLMs in practical applications is their inefficiency in handling long-text user behavior. Therefore, they propose the Behavior Aggregation Hierarchical Encoding (BAHE) to separate the encoding of user behavior from the interactions between behaviors, improving the efficiency of LLM-based click-through rate(CTR) modeling. 
\cite{2024_arXiv_LLM4SBR_LLM4SBR:A-Lightweight-and-Effective-Framework-for-Integrating} introduces a scalable two-stage LLM enhancement framework (LLM4SBR) tailored for session-based recommendation (SBR). It investigates the potential of integrating LLM with SBR models, prioritizing both effectiveness and efficiency. In scenarios involving short sequence data, LLM can deduce preferences directly using its language understanding capability, without the need for fine-tuning. This work is the first to propose an LLM enhancement framework for SBR. 

\subsection{Acceleration}
In the realm of LLM-based recommender systems, acceleration techniques are essential for optimizing performance and reducing latency. Given the substantial computational resources required by LLMs, improving efficiency in their deployment is critical. 
\cite{2024_arXiv_DARE_A-Decoding-Acceleration-Framework-for-Industrial-Deployable} identifies the inefficiency of knowledge generation when deploying LLM-based recommendation systems and proposes DARE. It is the first to integrate speculative decoding into LLM-based recommendations, thereby advancing the deployment of LLMs in recommendation systems. This work discovers two key traits of speculative decoding in recommendation systems and implements two enhancements: a Customized Retrieval Pool to improve retrieval efficiency, and Relaxed Verification to increase the number of accepted draft tokens. It has been deployed in online advertising scenarios within a large-scale commercial environment, achieving a 3.45x speedup while maintaining comparable downstream performance.

\subsection{Cold Start}

Cold start is one of the most challenging problems in recommender systems. Large Language Models possess vast world knowledge and can better understand the semantic information in product descriptions, as well as user preference information described in textual form. Therefore, incorporating LLMs has the potential to alleviate the cold start problem.
\cite{2023_CIKM_SR-Multi-Domain-Foundation_An-Unified-Search} proposes the S\&R Multi-Domain Foundation framework, which leverages LLMs to extract domain-invariant features and uses Aspect Gating Fusion to combine ID features, domain-invariant text features, and task-specific heterogeneous sparse features to obtain representations of queries and items. Additionally, samples from multiple search and recommendation scenarios are jointly trained with the Domain Adaptive Multi-Task module to create a multi-domain foundation model. They apply the S\&R Multi-Domain Foundation model to cold start scenarios using a pre-training and fine-tuning approach, achieving better performance than other state-of-the-art transfer learning methods.
Contemporary recommendation systems primarily rely on collaborative filtering techniques. However, this approach overlooks the substantial semantic information embedded in item text descriptions, resulting in suboptimal performance in cold start scenarios. \cite{2024_arXiv_LEARN_Knowledge-Adaptation-from} proposes the LLM-Driven Knowledge Adaptive Recommendation (LEARN) framework to effectively integrate the open-world knowledge encapsulated in LLMs into RS. This method significantly improves the revenue and AUC performance for cold start products. These enhancements are attributed to the robust representations generated by LEARN for products with sparse purchase histories.
\cite{2024_arXiv_LLM-KERec_Breaking-the-Barrier-Utilizing} proposes that traditional deep click-through rate (CTR) prediction models, which utilize feature interaction techniques through deep neural networks, have been widely applied in recommendation tasks. However, these models heavily rely on exposure samples and user feedback, limiting the performance of RS in cold start scenarios and making it difficult to handle the continuous emergence of new items. They are the first to leverage the reasoning capabilities of large language models as a medium to enhance scene preferences when recommending products to each user, achieving large-scale application of large language models in industrial scenarios.

\subsection{Dynamic Update}

Dynamic updates are essential for LLM-based recommender systems in industrial settings, as they ensure recommendations remain relevant and accurate by continually adapting to new user behaviors, content, and trends. Unlike static models that may quickly become outdated, dynamically updated models can respond in real-time or near real-time to changes in user interactions and preferences, which is critical in environments with rapidly changing data. This ongoing adaptability enhances user experience by providing timely and personalized recommendations.

In high-volume, fast-paced industrial applications, dynamic updates allow models to learn from streaming data, capturing the most recent user interactions and evolving preferences. This capability not only improves recommendation relevance but also enables businesses to quickly adapt to shifts in user behavior, maintaining competitiveness and optimizing user engagement.

Due to the continuous influx of new content and products every minute, temporal information is crucial for understanding user preferences and the evolution of user-item interactions over time. The integration of large language models in dynamic recommender systems remains largely unexplored, primarily due to the complexity of adapting LLMs to predict dynamically changing data. For the first time, \cite{2024_arXiv_DynLLM_When-Large-Language-Models} leverages the continuous time dynamic graphs framework to integrate LLMs with dynamic recommendation. This integration offers a novel perspective for dynamically modeling user preferences and user-item interactions. They introduce a novel LLM-enhanced dynamic recommendation task based on continuous time dynamic graphs, and propose the DynLLM model, which effectively integrates LLM-enhanced information with temporal graph data.
\cite{2024_arXiv_HSTU_Actions-speak-louder} proposes HSTU architecture, designed for high cardinality, non-stationary streaming recommendation data. They first convert ranking and retrieval tasks into serialised prediction tasks and propose pointwise aggregated attention to improve the original Transformer model.It also contains solutions optimised for accelerated inference and reduced memory usage. The solution has been deployed on a large internet platform with billions of users. 
\cite{2024_SIGMOD-Companion_COSMO} presents COSMO, a scalable system designed to extract user-centric commonsense knowledge from extensive behavioral data and construct industry-scale knowledge graphs, thereby enhancing a variety of online services. Finally, COSMO has been deployed in various Amazon search applications. The deployment centers around an efficient feature store and asynchronous cache store, ensuring streamlined processing and cost-effective management of customer queries and model responses.

\subsection{Business Customization Requirements}

In industrial applications, recommender systems need to be tailored to meet the unique requirements of different businesses, each with distinct user bases, content types, and objectives. A one-size-fits-all approach is insufficient. For instance, e-commerce platforms require systems that handle diverse products, seasonal trends, and dynamic pricing, while media streaming services prioritize content consumption patterns and viewer engagement. Customizable models that adapt to these specific contexts, user behaviors, and domain knowledge are essential to optimize performance and achieve business-specific goals, such as increasing engagement, conversion rates, or retention.

Recent advancements in LLM-based recommender systems demonstrate significant potential for such customization. By fine-tuning these models to understand specific user behaviors and integrate domain-specific and external knowledge, businesses can create more adaptable and context-aware systems. This flexibility allows recommendation engines to deliver more relevant and effective suggestions, enhancing user experiences and better aligning with business objectives.

\cite{2024_WWW_SINGLE_Modeling-User-Viewing-Flow} proposes a method called SINGLE for user view stream modeling in article recommendation tasks, which includes two parts: Constant View Stream Modeling and Instantaneous View Stream Modeling.
First, they use LLMs to capture constant user preferences from previously clicked articles. Then, they model the instantaneous view stream of users by leveraging the interaction between the user's article click history and the candidate articles.
\cite{2024_arXiv_Ad-Recommendation} proposes an industrial advertising recommendation system that addresses sequential features, numerical features, pre-trained embedding features, and sparse ID features. Additionally, they proposes methods to effectively tackle two key challenges related to feature representation: embedding dimensional collapse and interest entanglement across various tasks or scenarios. They explore several training techniques to facilitate model optimization, reduce bias, and enhance exploration. Furthermore, they introduce three analytical tools that allow for a comprehensive investigation of feature correlations, dimensional collapse, and interest entanglement. 
\cite{2024_WWW_NoteLLM_a-re} proposes a new unified framework, NoteLLM, which utilizes LLMs to address the I2I recommendation problem for notes. They use a note compression prompt to compress a note into a unique word and further learn embeddings of potentially related notes through a contrastive learning approach. 
Recommendation systems need to combine semantic information with behavioral data. The current mainstream approach is to use user and item ID embeddings to enhance recommendation performance, but these embeddings often fail to capture the content relevance of the items themselves, particularly in cold start problems and similarity-based recommendation scenarios. 
\cite{2023_RecSys_Beyond-Labels-Leveraging} discusses the importance of content metadata in movie recommendation systems, particularly the role of genre labels in understanding user preferences and providing personalized recommendations. They point out challenges associated with using genre labels, such as inconsistent genre definitions, the subjectivity of genre labels, the presence of mixed genres, and the inability of genre labels to capture the intensity or degree of genres within videos.
\cite{2024_arXiv_TRAWL_External-Knowledge-Enhanced} introduced a recommendation system method called TRAWL (External Knowledge-Enhanced Recommendation with LLM Assistance). TRAWL leverages large language models to extract recommendation-relevant knowledge from raw external data and uses a contrastive learning strategy for adapter training to enhance behavior-based recommendation systems.
\cite{2023_RecSys_HKFR_hetero} emphasizes the importance of analyzing and mining user heterogeneous behaviors in recommendation systems. They propose HKFR, which extracts and integrates heterogeneous knowledge from user behaviors using LLMs to achieve personalized recommendations. By performing instruction tuning on LLMs and combining heterogeneous knowledge with recommendation tasks, they significantly improve recommendation performance.
\cite{2024_SIGMOD-Companion_COSMO} proposes a fine-tuning language model(COSMO-LM) on a curated set of e-commerce annotated data, structured as instructions, to produce high-quality commonsense knowledge that aligns with human preferences. To obtain large-scale and diverse instructional data, an automated instruction generation pipeline is developed, leveraging massive user behaviors. By expanding product domains, relation types, and fine-tuning tasks, the approach enables scalable knowledge extraction.

\section{Challenges and Opportunities}
\label{sec_Challenges and Opportunities}

% --------------- figure: Challenge ---------------
\begin{figure*}
    \centering
    \includegraphics[width=0.7\linewidth]{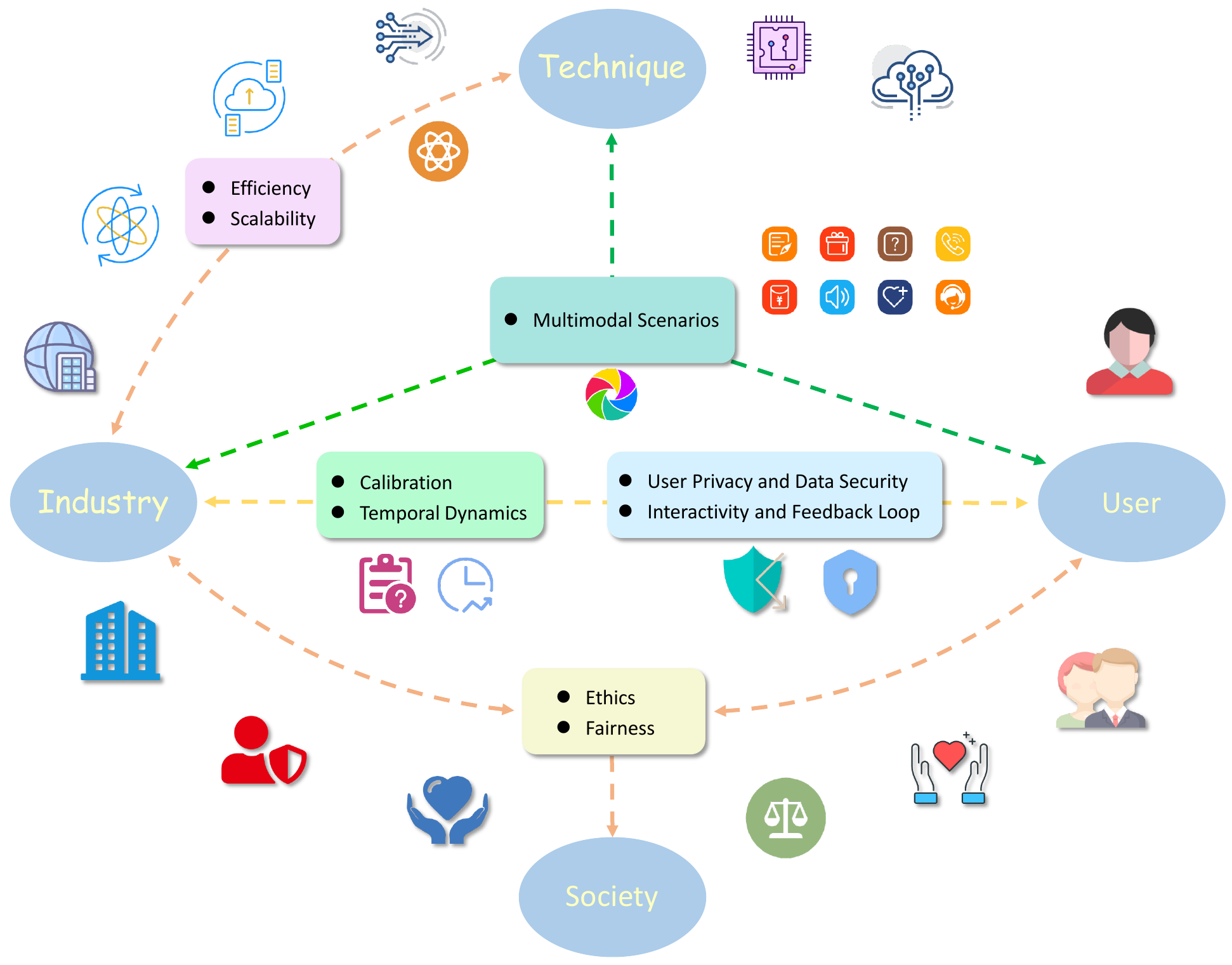}
    \caption{The challenges faced by LLM-based recommender system.}
    \label{fig:challenges}
\end{figure*}

The integration of large language models (LLMs) into recommender systems offer the potential to revolutionize how recommendations are generated, leveraging vast amounts of data and complex contextual understanding to provide users with highly tailored suggestions. However, this evolution comes with its own set of challenges that need careful consideration. In this section, we discuss the challenges and opportunities. Fig. \ref{fig:challenges} illustrates the relationships between various challenges and their impact on users, industry, technology, and society.

\subsection{Calibration}
In LLM-based Generative Recommender Systems, recommendations are generated directly based on user input or context, but the strength of user preferences can vary significantly. For example, one user might have a top 10 list with high preference scores (e.g., 0.9, 0.89), while another’s might have much lower scores (e.g., 0.6, 0.2). Although both users receive recommendations they like, the inconsistency in preference strength poses challenges, especially in business scenarios like ad placements or engagement predictions, where precise preference understanding is crucial. Addressing this requires integrating a calibration mechanism to adjust generated recommendations based on the relative strength of user preferences. This could involve normalizing preference scores or considering variations across users, ensuring not just relevant recommendations but also a more accurate reflection of a user’s true interest.

\subsection{Temporal Dynamics}
Temporal dynamics enable LLM-based recommender systems to adapt to changing user preferences and behaviors. As interests shift with trends and events, advanced temporal modeling is necessary. Recommendations should balance recent interactions with historical data, employing weighting and decay strategies. Seasonal variations also require contextual adaptation. Strategies like time-aware embeddings, temporal attention mechanisms, and real-time data processing enhance personalization and deliver timely, relevant recommendations, boosting user satisfaction and engagement.

\subsection{Scalability}
Scalability in LLM-based recommender systems enables efficient handling of increasing data volumes and user interactions. \textit{Horizontal scalability} expands capacity through additional servers and load balancing (e.g., Apache Spark, Kubernetes), while \textit{vertical scalability} enhances server performance. \textit{Architectural scalability} employs modular designs and containerization (e.g., Docker) for flexible deployment. \textit{Data scalability} utilizes NoSQL databases and distributed file systems (e.g., Apache Hadoop) to manage large datasets. Strategies include combining scaling approaches, adopting modular architectures, and implementing robust data management solutions, with monitoring and auto-scaling mechanisms ensuring consistent performance and reliability.

\subsection{Efficiency}
Efficiency in LLM-based recommender systems involves optimizing performance while managing computational resources and costs. \textit{}{Computational Efficiency} focuses on reducing latency and improving responsiveness by optimizing model architectures and using hardware accelerators such as GPUs or TPUs. \textit{Operational Efficiency} involves resource and cost management through strategies such as workload distribution and cost-effective infrastructure (e.g., cloud computing). \textit{Energy Efficiency} addresses the environmental impact by employing techniques such as model compression and quantization. Key strategies for efficiency include optimizing model performance, leveraging hardware acceleration, implementing efficient inference methods, managing resources effectively, and adopting energy-efficient practices. These approaches ensure that the system operates efficiently and sustainably while meeting user demands.

\subsection{Multimodal Recommendation Scenarios}
LLM-based multimodal recommender systems enhance personalization by integrating diverse data types, such as text, images, audio, and video. By leveraging LLMs, these systems achieve a comprehensive understanding of user preferences, improving the relevance and accuracy of recommendations. For example, in e-commerce, combining product descriptions, images, and customer reviews through LLMs offers deeper insights into user preferences. Despite challenges in data fusion and computational complexity, LLMs support effective cross-modal representation learning, enhancing recommendation diversity and user engagement by delivering contextually aware content across various domains.

\subsection{User Privacy and Data Security}
LLM-based recommender systems require extensive user data for personalization, necessitating strong security measures. Compliance with data protection regulations involves transparent practices, explicit user consent, and user control over personal data. Techniques like anonymization and encryption safeguard personally identifiable information (PII) during storage and transmission. Access control mechanisms like RBAC and MFA prevent unauthorized access, with regular audits to identify vulnerabilities. Strategies such as regulatory compliance, data minimization, and user education work together to enhance data security and foster user trust.

\subsection{Interactivity and Feedback Loop}
Interactivity and feedback loops enhance user engagement in LLM-based recommender systems. Users can actively adjust settings and customize preferences, tailoring their experiences. Feedback loops capture user input to refine algorithms and adapt recommendations to changing preferences, fostering trust. Key strategies include customizable settings, intuitive interfaces, real-time personalization, structured feedback integration, and engagement incentives. These approaches create responsive, user-centric systems that evolve with user input, improving satisfaction and trust.

\subsection{Ethics}
Ethics in LLM-based recommender systems ensure fair, transparent, and responsible operation. Transparency through explainable AI (XAI) helps users understand recommendations. Compliance with standards like GDPR protects privacy and data security. Fairness involves auditing biases and using diverse datasets to prevent discrimination. Content filtering and moderation are vital to avoid harmful material. Continuous monitoring ensures accountability and adapts to new ethical challenges, including the system's impact on mental health. Regular updates and stakeholder engagement, including input from users and ethicists, help maintain ethical standards and build trust, enabling responsible operation and positive societal impact.

\subsection{Fairness}
Fairness in LLM-based recommender systems ensures equal treatment across user groups and addresses biases related to race, gender, and age. LLMs must provide equitable recommendations by analyzing and mitigating biases. Systems should avoid favoring popular content to prevent biased exposure. Techniques like re-balancing datasets and applying fairness constraints are essential. Additionally, considering diverse user feedback helps reduce bias. Strategies include using fairness metrics and conducting regular audits to adapt to changing behaviors, fostering positive user experiences and supporting ethical practices.

% --------------- figure: statistics of collected papers ---------------
\begin{figure*}
    \centering
    \begin{minipage}[t]{0.41\linewidth}
        \centering
        \includegraphics[width=\linewidth]{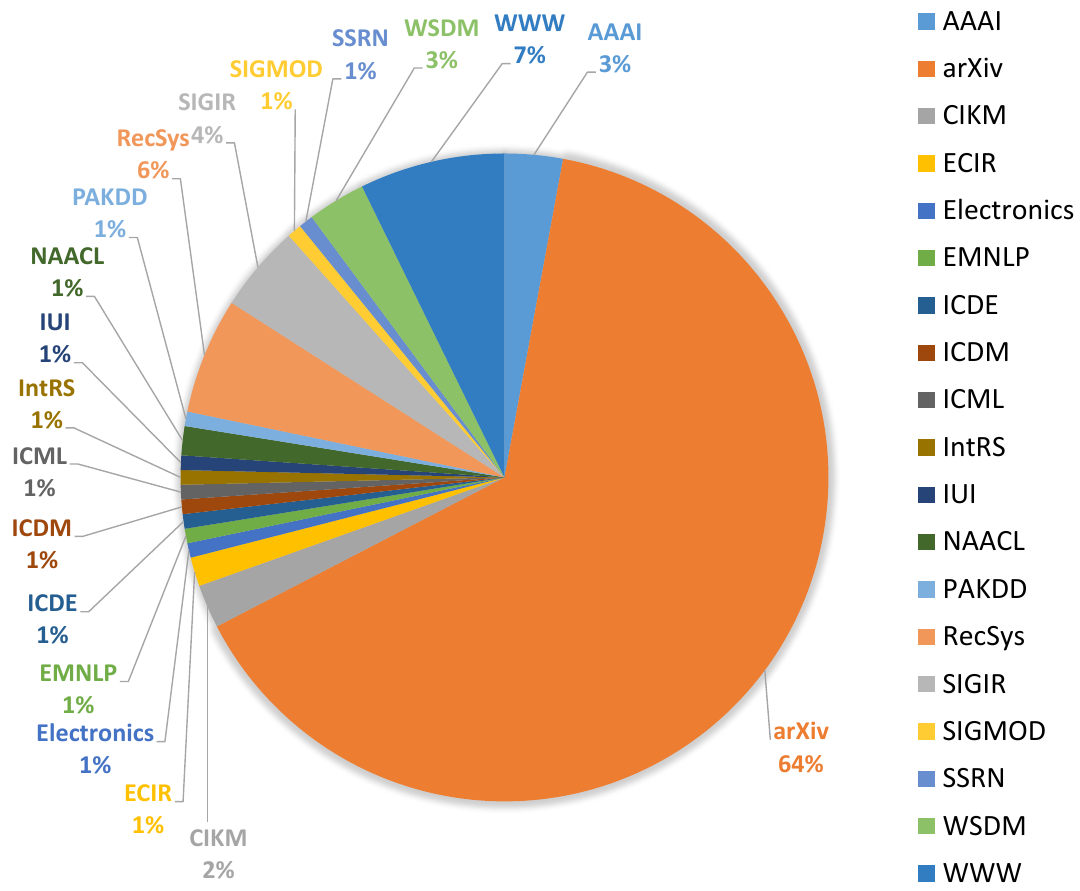}
        \caption*{(a) Distribution of cited papers across academic conferences and journals.}
        \label{fig:statistics_of_collected_papers_1}
    \end{minipage}
    % \hfill
    \begin{minipage}[t]{0.48\linewidth}
        \centering
        \includegraphics[width=\linewidth]{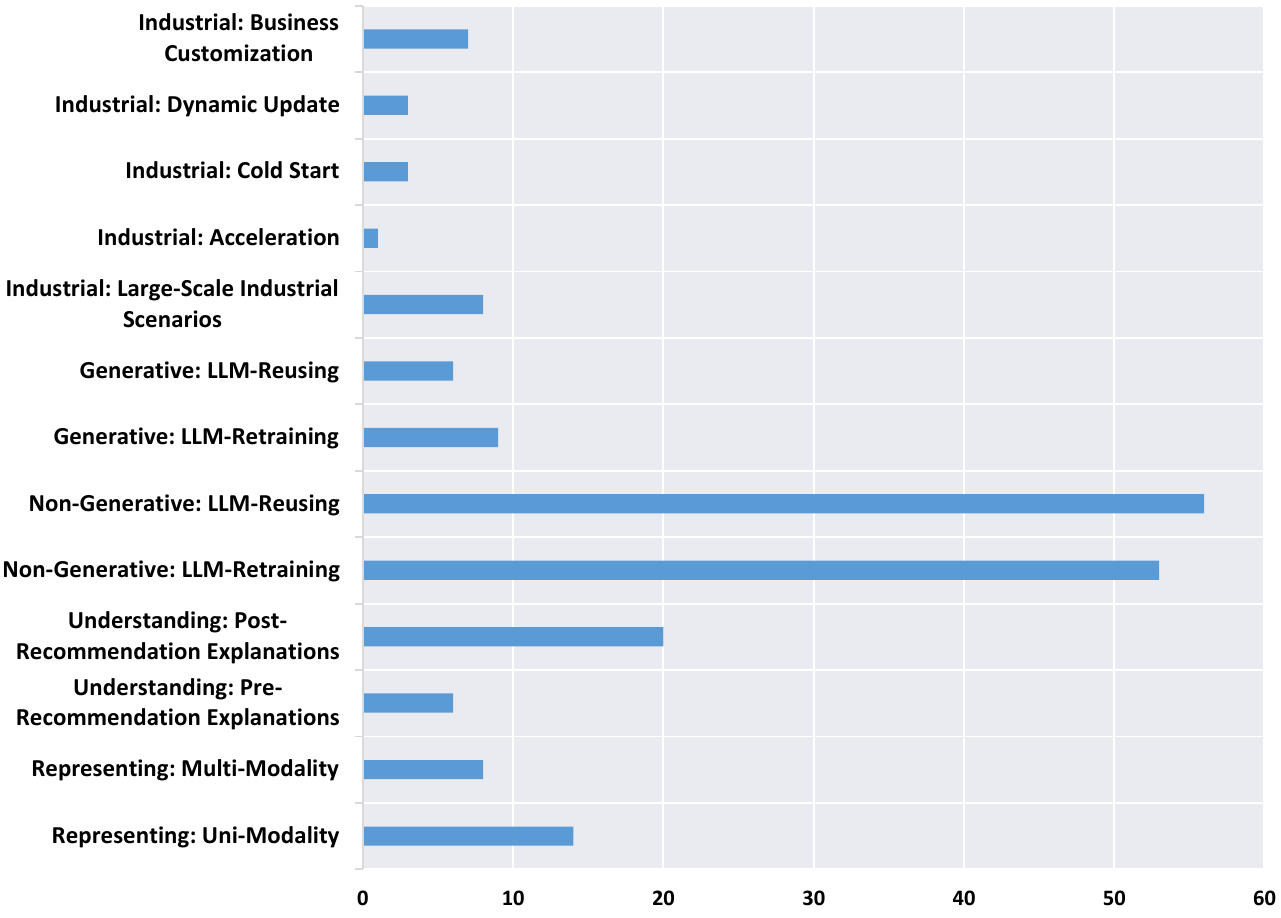}
        \caption*{(b) Count of cited papers in different sections.}
        \label{fig:statistics_of_collected_papers_2}
    \end{minipage}
    \caption{Statistics of papers on LLM-based recommender systems collected in this survey.}
    \label{fig:statistics_of_collected_papers_combined}
\end{figure*}

\section{Conclusion}
\label{sec_Conclusion}

This paper presents a comprehensive review that not only delineates recent advances in the field, but also discusses the challenges faced by LLM-based recommender systems. 
Specifically, a novel taxonomy is introduced, which provides a structured approach to understanding the integration of LLMs in recommender systems and their deployment in industry. This taxonomy is organized into a three-tier framework that encapsulates the progression from theoretical research to practical application. Each tier is designed to build upon the previous, with Representing and Understanding enhancing the capabilities for Scheming and Utilizing, which in turn facilitates the Industrial Deployment of recommender systems. Unlike existing surveys, this work offers an original perspective by constructing a taxonomy inspired by insights from the recommendation community rather than adhering solely to the established LLM categorizations. We summarize the articles collected on LLM-based recommender systems from various conferences and journals as presented in Fig. \ref{fig:statistics_of_collected_papers_combined}. We observe an intriguing trend that, despite the emphasis on generative recommendation by the majority of papers on LLM-based recommender systems, most studies still adhere to conventional recommendation processes, utilizing LLMs as technical components. Future research should delve into the superior text comprehension and generation capabilities of LLMs to unlock further potential for authentic generative recommendation. Future work will enhance technical performance, ethical standards, and novel methods to boost the efficacy and security of LLM-based recommender systems, making them more adaptive and beneficial to users and society.


% \begin{acks}
% This work is supported by the Youth Fund of the National Natural Science Foundation of China (No. 62206107 and No. 62406127) and the National Science and Technology Major Project under Grant No. 2023YFF0905400.
% \end{acks}

\section*{Disclaimer}
This survey explores the current state and future directions of Large Language Models (LLMs) in recommender systems. The findings and viewpoints are independent contributions from the authors, particularly regarding ethics and biases. As LLM technologies evolve, limitations may improve, but new challenges may arise. We encourage readers to view this survey as a starting point for future research and practical validation. Ongoing developments in LLM-based recommender systems may not be fully covered, and we welcome constructive feedback and suggestions.

% \bibliographystyle{ACM-Reference-Format}
\bibliographystyle{plainnat}
\bibliography{Survey_arXiv}

% \appendix

\end{document}